\documentclass{article}
\pdfoutput=1
\usepackage{arxiv}
\usepackage{setspace}
\usepackage[utf8]{inputenc} 
\usepackage[T1]{fontenc}    
\usepackage[hyperfootnotes=false]{hyperref}
\usepackage{url}            
\usepackage{booktabs}       
\usepackage{amsfonts}       
\usepackage{nicefrac}       
\usepackage{microtype}      
\usepackage{lipsum}
\usepackage{xcolor}
\usepackage{amsmath}
\usepackage{rotating}
\usepackage{multirow}
\usepackage{lineno}
\usepackage{xstring}
\usepackage{threeparttable}
\usepackage[hang,flushmargin,multiple]{footmisc}
\usepackage[round,authoryear]{natbib}
\usepackage[singlelinecheck=true,labelfont=bf]{caption}
\usepackage{tikz}
\usetikzlibrary{decorations.pathreplacing,shapes.arrows,chains,matrix,positioning}
\usepackage{subcaption}
\usepackage{appendix}

\definecolor{awesome}{rgb}{1.0, 0.13, 0.32}
\definecolor{pear}{rgb}{0.82, 0.89, 0.19}
\definecolor{pastelyellow}{rgb}{0.99, 0.99, 0.59}
\definecolor{fluor}{rgb}{0.8, 1.0, 0.0}
\definecolor{cinnamon}{rgb}{0.82, 0.41, 0.12}

\title{Macroeconomic forecasting with LSTM and mixed frequency time series data}

\author{
  Sarun Kamolthip\\
  Faculty of Economics, Khon Kaen University\\
  Khon Kaen 40002, Thailand \\
  \texttt{saruka@kku.ac.th}
}

\begin{document}
\maketitle

\begin{abstract}
This paper demonstrates the potentials of the long short-term memory (LSTM) when applying with macroeconomic time series data sampled at different frequencies. We first present how the conventional LSTM model can be adapted to the time series observed at mixed frequencies when the same mismatch ratio is applied for all pairs of low-frequency output and higher-frequency variable. To generalize the LSTM to the case of multiple mismatch ratios, we adopt the unrestricted Mixed DAta Sampling (U-MIDAS) scheme \citep{foroni2015unrestricted} into the LSTM architecture. We assess via both Monte Carlo simulations and empirical application the out-of-sample predictive performance. Our proposed models outperform the restricted MIDAS model even in a set up favorable to the MIDAS estimator. For real world application, we study forecasting a quarterly growth rate of Thai real GDP using a vast array of macroeconomic indicators both quarterly and monthly. Our LSTM with U-MIDAS scheme easily beats the simple benchmark AR(1) model at all horizons, but outperforms the strong benchmark univariate LSTM only at one and six months ahead. Nonetheless, we find that our proposed model could be very helpful in the period of large economic downturns for short-term forecast. Simulation and empirical results seem to support the use of our proposed LSTM with U-MIDAS scheme to nowcasting application. 
\end{abstract}

\newpage
\section{Introduction}
Forecasting the present or very near future (also called “nowcasting”) of an economic activity in order to update the state of economy is one of the crucial questions in macroeconomic area \citep{giannone2008nowcasting}. Nonetheless, the key macroeconomic data periodically released by government agencies, for example, a quarterly Gross Domestic Product (QGDP), are typically only available after several weeks or months and often revised a few months later (see \cite{ghysels2018forecasting} for a recent discussion of macroeconomic data revision and publication delays). The low-frequency nature of QGDP data is challenging to policy makers and those who are in need of updating the state of economy for their decision making. This is so called mixed frequency data problem.

Researchers and practitioners working with the mixed frequency time series data usually pre-filter the data so that the frequency of left-hand and right-hand variables are equal. This raises the problem of how to exploit potentially useful information contained in the higher-frequency time series data. Various approaches have been proposed so far in the literature to avoid pre-filtering and to benefit from the information contained in the time series data observed at different frequencies (see \cite{foroni2013survey} for a comprehensive survey on the methods when dealing with mixed frequency data). Mixed DAta Sampling, denoted MIDAS, approach was first introduced by \cite{ghysels2006predicting, ghysels2007midas} and has since been one of the main strands when modelling mixed frequency time series. MIDAS uses a flexible functional form to parameterize the weight given to each lagged regressors, such that the regressand and regressors can be sampled at different frequencies. Apart from capturing the rich information contained in high-frequency data, MIDAS also avoids the potential problems of information loss, measurement error, and timeliness caused by frequency transformation. Notable studies focusing on macroeconomic indicators include \cite{clements2008macroeconomic, armesto2009measuring, marcellino2010factor, kuzin2011midas, breitung2015forecasting}. As long as the differences in sampling frequencies are large, implying that a curse of dimensionality might be relevant, MIDAS can be expected to perform well as the use of distributed lag polynomials can help avoiding parameter proliferation. However, when the differences in sampling frequencies are small, \cite{foroni2015unrestricted} show that the unrestricted lag polynomials in MIDAS regressions (also called U-MIDAS) generally performs better than the restricted MIDAS. 
\cite{koenig2003use} is one of the first papers which apply the U-MIDAS to forecast quarterly changes in real GDP. \cite{clements2008macroeconomic} also consider U-MIDAS in their empirical forecast comparisons but call these U-MIDAS-like models as \textit{mixed-frequency distributed lag} (MF-DL) models. Nonetheless, \cite{foroni2013survey} is the first work that provides the theoretical framework and parameter identification conditions for the underlying high-frequency model in the U-MIDAS regressions.

Several sources of nonstandard high-frequency data from private companies such as Google and Twitter are available for short-term economic prediction \citep{ettredge2005using,vosen2011forecasting,choi2012predicting,d2017predictive}. Nonetheless, most of the nowcasting applications of MIDAS literature still rely on the standard high-frequency macroeconomic data (e.g. industrial production, employment, and capacity utilization) usually released by government agencies \citep{clements2008macroeconomic,marcellino2010factor,kuzin2011midas}. An important reason is the usually limited effective sample size of low-frequency data when involving a large number of variables sampled at high-frequency. If, for example, we consider to use 30 variables at 10 lags to forecast the quarterly GDP growth, MIDAS regression model with two parameters lag polynomial needs to estimate 60 parameters, whereas U-MIDAS regression model needs to estimate 300 parameters. However, the GDP growth series of developed countries are rarely, if any, more than 300 observations and far less than that for the developing countries.

In order to deal with such high-dimensional data environment, recent MIDAS literature start to rely on machine learning models such as tree-based algorithms \citep{qiu2020forecasting} and sparse-group LASSO (sg-LASSO) regression \citep{babii2021machine}. To my knowledge, \cite{xu2019artificial} is the only work that applies deep learning architecture with mixed frequency data. They develop the ANN-MIDAS and ANN-U-MIDAS models which integrate the MIDAS projection approach into the artificial neural networks (ANNs). The proposed models outperform the benchmark MIDAS and U-MIDAS models in both Monte Carlo simulations and empirical applications. They also find that the ANN-MIDAS model is superior to ANN-U-MIDAS in terms of goodness-of-fit and predictive ability. So far, the potentials of machine learning and deep learning models on mixed frequency data have barely been explored. This paper fills these gaps in the literature demonstrating the use of LSTM with with time series data sampled at different frequencies.

We presents the potentials of deep learning models when applying with mixed frequency macroeconomic time series data. We apply the long short-term memory networks (LSTM) introduced by \cite{hochreiter1997long}. The LSTM is developed to overcome the vanishing error problems found in standard recurrent neural networks (RNNs) when many time lags between relevant input events and target signals are present (see \cite{hochreiter2001gradient} for more discussion). The LSTM has been one of the main workhorses in natural language processing, language translation, speech recognition, and sentiment analysis. LSTM and its variations has recently gained lots of attentions for financial time series forecasting application. For a comprehensive survey on financial time series forecasting with deep learning algorithms see \cite{sezer2020financial}. To deal with data sampled at different frequencies, we rely on the U-MIDAS scheme. From machine learning perspective, the U-MIDAS scheme serves as an alternative feature engineering process that transforms the higher-frequency variables into a low-frequency vector, denoted frequency alignment, and allows us to exploit potentially useful information contained in the high-frequency time series data. We first assess via Monte Carlo simulation the predictive performance of our proposed LSTM based models, and find that our nowcasts and forecasts are superior to those produced by the benchmark models. For real world application, we study both nowcasting (i.e. 1-, -2, and 3-month ahead forecasts) and forecasting (i.e. 6-, 9-, and 12-month ahead forecasts) a quarterly growth rate of Thai real GDP using a vast array of macroeconomic indicators both quarterly and monthly. Although our proposed model outperform the benchmark models only at one and six month ahead, we find that our proposed model could be very helpful in the period of large economic downturns. We additionally compare our forecasts with those posted by the Bank of Thailand (hereinafter BOT). Our annualized QGDP forecasts are superior to the BOT implied annual GDP forecasts at shorter horizon (i.e. 1- and 2-month ahead).

The remainder of this paper is organized as follows: Section 2 presents alternative methods for the LSTM architecture when applying time series data sampled at different frequencies. Section 3 reports on Monte Carlo simulations the predictive performance to illustrate the potentials of the proposed model when compared to the original MIDAS regression. Section 4 provides empirical evidence, and we conclude in Section 5.


\section{Alternative methods to apply mixed frequency time series data with LSTM}
This section presents how the time series data sampled at mixed frequency can be accommodated into the LSTM architectures in both conventional and unconventional fashions. We first discuss the basic setup of LSTM and how we can apply it when there is only one frequency mismatch between output variable and all variables in the model specification. To deal with multiple frequency mismatches, we then briefly introduce the MIDAS approach and discuss how we can adopt the unrestricted MIDAS scheme into the LSTM architectures. 

The LSTM is capable of learning the complex long time lag tasks without loss of short time lag capabilities by enforcing constant error flow through the memory cells, a unit of computation that replace traditional artificial neurons in the hidden layer of a network. However, the original version of LSTM does not work well when dealing with a continuous input stream \citep{gers2000learning}. This paper apply the LSTM with a \textit{forget gate} introduced by \cite{gers2000learning}. Following the description of \cite{graves2013generating} and \cite{fischer2018deep}, the LSTM is basically composed of an input layer, one or more hidden layers, and an output layer. The hidden layer(s) consists of a set of recurrently connected memory blocks. Each block contains one or more self-connected memory cells and three multiplicative units—an input gate ($i_t$), a forget gate ($f_t$), and an output gate ($o_t$)—that provide continuous analogues of write, read and reset operations for the cell state ($c_t$), respectively. For the LSTM used in this paper \citep{gers2000learning}, the hidden layer function is implemented by the following group of functions:
\begin{linenomath}
\begin{subequations}\label{eq:lstm}
    \begin{align}
    i_t &= \sigma(W_{xi}x_t + W_{ti}h_{t-1} + W_{ci}c_{t-1} + b_i) \\
    f_t &= \sigma(W_{xf}x_t + W_{hf}h_{t-1} + W_{cf}c_{t-1} + b_f) \\
    c_t &= f_{t}c_{t-1} + i_{t}tanh(W_{xc}x_t + W_{hc}h_{t-1} + b_c) \\
    o_t &= \sigma(W_{xo}x_t + W_{ho}h_{t-1} + W_{c0}c_t + b_0 \\
    h_t &= o_t tanh(c_t)
    \end{align}
\end{subequations}
\end{linenomath}
where $\sigma$ is the logistic sigmoid function; and the input gate, forget gate, output gate, and cell state activation vectors are all the same size as the hidden vector ($h_t$). The weight matrices $Wx_i$,$W_{hi}$,$W_{ci}$,$W_{xf}$,$W_{hf}$,$W_{cf}$,$W_{xc}$,$W_{hc}$,$W_{xo}$,$W_{ho}$, and $W_{co}$ have the obvious meaning, for example, $W_{hi}$ is the hidden-input gate matrix, $W_{xo}$ is the input-output gate matrix. The $b$ terms denote bias vectors (e.g. $b_i$ is input bias vector). See \cite{fischer2018deep} for more detailed description of working sequences of LSTM.

\subsection{Single frequency mismatch}
\label{sub:single}

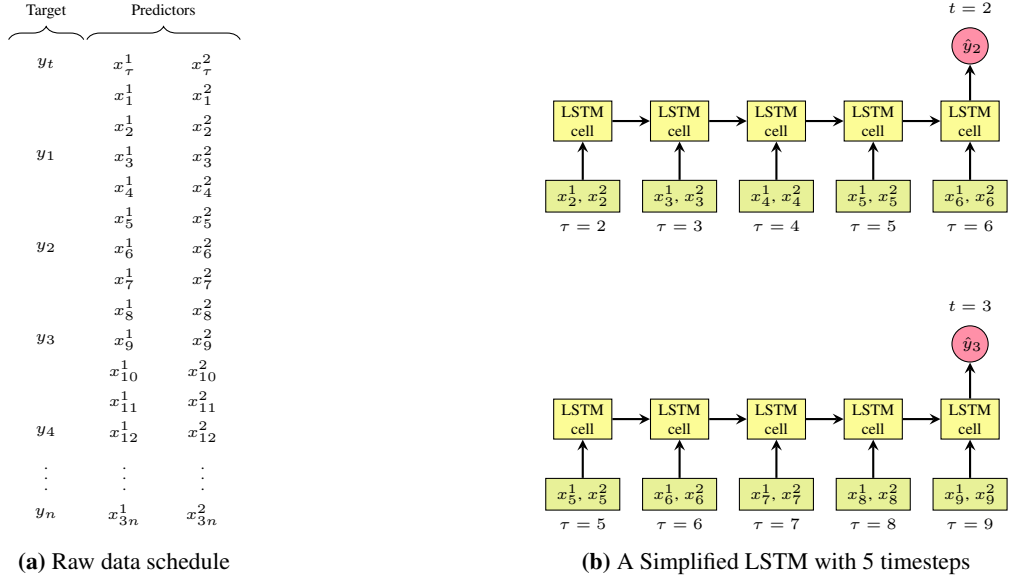
\begin{figure}[t]
    \centering
    \begin{subfigure}[t]{0.4\linewidth}
    \centering
    \tiny
    \begin{tikzpicture}[cell/.style={rectangle},
    row 1/.style={nodes={draw=none,fill=none}},
    space/.style={minimum height=2em,matrix of nodes,row sep=-\pgflinewidth,column sep=-\pgflinewidth, column 1/.style={nodes={anchor=west,minimum width=5em}}},
    B/.style = {decorate,decoration={brace,amplitude=5pt,pre=moveto,pre length=1pt,post=moveto,post length=1pt,raise=1mm}},
    text depth=0.5ex,nodes in empty cells]
    \matrix (second) [
    space,
    column 2/.style={anchor=west,nodes={cell,minimum width=5em}},
    column 3/.style={anchor=west,nodes={cell,minimum width=5em}},
    ]{
    $y_{t}$   & $x_{\tau}^{1}$  & $x_{\tau}^{2}$ \\
              & $x_{1}^{1}$  & $x_{1}^{2}$ \\
              & $x_{2}^{1}$  & $x_{2}^{2}$ \\
    $y_{1}$   & $x_{3}^{1}$  & $x_{3}^{2}$ \\
              & $x_{4}^{1}$  & $x_{4}^{2}$ \\
              & $x_{5}^{1}$  & $x_{5}^{2}$ \\
    $y_{2}$   & $x_{6}^{1}$  & $x_{6}^{2}$ \\
              & $x_{7}^{1}$  & $x_{7}^{2}$ \\
              & $x_{8}^{1}$  & $x_{8}^{2}$ \\
    $y_{3}$   & $x_{9}^{1}$  & $x_{9}^{2}$ \\
              & $x_{10}^{1}$ & $x_{10}^{2}$ \\
              & $x_{11}^{1}$ & $x_{11}^{2}$ \\
    $y_{4}$   & $x_{12}^{1}$ & $x_{12}^{2}$ \\
    \vdots    & \vdots       & \vdots \\
    $y_{n}$   & $x_{3n}^{1}$ & $x_{3n}^{2}$ \\
    };
    \draw[B] (second-1-2.north west) -- node[above=3mm] {Predictors} (second-1-3.north east);
    \draw[B] (second-1-1.north west) -- node[above=3mm] {Target} (second-1-1.north east);

    \end{tikzpicture}
    \caption{Raw data schedule}
    \label{fig:fig1a}
    \end{subfigure}
    \hfill
    \begin{subfigure}[t]{0.55\linewidth}
    \centering
    \tikzstyle{input}=[rectangle,draw,fill=pear!50,text width=4em,minimum height=2em]
    \tikzstyle{lstm}=[rectangle,draw,fill=pastelyellow,text width=3em,minimum height=1em]
    \tikzstyle{output}=[circle,draw,fill=awesome!50,text width=1em,minimum height=2em]
    \tikzstyle{truey}=[circle,draw,fill=cinnamon!50,text width=1em,minimum height=2em]
    \tikzstyle{arrow} = [thick,->,>=stealth]
    \tiny
    \begin{tikzpicture}[text centered,node distance=1cm and 0.5cm,auto]
    \node [output] (output1) {$\hat{y}_2$};
    \node [lstm,below of=output1] (cell51) {LSTM cell};
    \node [lstm,left =of cell51] (cell41) {LSTM cell};
    \node [lstm,left =of cell41] (cell31) {LSTM cell};
    \node [lstm,left =of cell31] (cell21) {LSTM cell};
    \node [lstm,left =of cell21] (cell11) {LSTM cell};
    \node [input,below of= cell51] (input61) {$x_{6}^1$, $x_{6}^2$};
    \node [input,below of= cell41] (input51) {$x_{5}^1$, $x_{5}^2$};
    \node [input,below of= cell31] (input41) {$x_{4}^1$, $x_{4}^2$};
    \node [input,below of= cell21] (input31) {$x_{3}^1$, $x_{3}^2$};
    \node [input,below of= cell11] (input21) {$x_{2}^1$, $x_{2}^2$};
    \node [draw=none,below of= input21,yshift=0.6cm] (t2) {$\tau=2$};
    \node [draw=none,below of= input31,yshift=0.6cm] (t3) {$\tau=3$};
    \node [draw=none,below of= input41,yshift=0.6cm] (t4) {$\tau=4$};
    \node [draw=none,below of= input51,yshift=0.6cm] (t5) {$\tau=5$};
    \node [draw=none,below of= input61,yshift=0.6cm] (t6) {$\tau=6$};
    \node [draw=none,above of= output1,yshift=-0.5cm] (o1) {$t=2$};
    \draw [arrow] (input61) -- (cell51);
    \draw [arrow] (input51) -- (cell41);
    \draw [arrow] (input41) -- (cell31);
    \draw [arrow] (input31) -- (cell21);
    \draw [arrow] (input21) -- (cell11);
    \draw [arrow] (cell11) -- (cell21);
    \draw [arrow] (cell21) -- (cell31);
    \draw [arrow] (cell31) -- (cell41);
    \draw [arrow] (cell41) -- (cell51);
    \draw [arrow] (cell51) -- (output1);

    \node [output,below =of input61,yshift=-0.5cm] (output2) {$\hat{y}_3$};
    \node [lstm,below of=output2] (cell52) {LSTM cell};
    \node [lstm,left =of cell52] (cell42) {LSTM cell};
    \node [lstm,left =of cell42] (cell32) {LSTM cell};
    \node [lstm,left =of cell32] (cell22) {LSTM cell};
    \node [lstm,left =of cell22] (cell12) {LSTM cell};
    \node [input,below of= cell52] (input62) {$x_{9}^1$, $x_{9}^2$};
    \node [input,below of= cell42] (input52) {$x_{8}^1$, $x_{8}^2$};
    \node [input,below of= cell32] (input42) {$x_{7}^1$, $x_{7}^2$};
    \node [input,below of= cell22] (input32) {$x_{6}^1$, $x_{6}^2$};
    \node [input,below of= cell12] (input22) {$x_{5}^1$, $x_{5}^2$};
    \node [draw=none,below of= input22,yshift=0.6cm] (t22) {$\tau=5$};
    \node [draw=none,below of= input32,yshift=0.6cm] (t32) {$\tau=6$};
    \node [draw=none,below of= input42,yshift=0.6cm] (t42) {$\tau=7$};
    \node [draw=none,below of= input52,yshift=0.6cm] (t52) {$\tau=8$};
    \node [draw=none,below of= input62,yshift=0.6cm] (t62) {$\tau=9$};
    \node [draw=none,above of= output2,yshift=-0.5cm] (o2) {$t=3$};
    \draw [arrow] (input62) -- (cell52);
    \draw [arrow] (input52) -- (cell42);
    \draw [arrow] (input42) -- (cell32);
    \draw [arrow] (input32) -- (cell22);
    \draw [arrow] (input22) -- (cell12);
    \draw [arrow] (cell12) -- (cell22);
    \draw [arrow] (cell22) -- (cell32);
    \draw [arrow] (cell32) -- (cell42);
    \draw [arrow] (cell42) -- (cell52);
    \draw [arrow] (cell52) -- (output2);
    \end{tikzpicture}
    \caption{A Simplified LSTM with 5 timesteps}
    \label{fig:fig1b}
    \end{subfigure}
\caption{Mixed frequency time series with single frequency mismatch ratios. (a) presents a raw data schedule in which the raw time series data are aligned to match with their actual period of occurrence. (b) presents the first two feasible sequences of LSTM when timesteps is set to 5.}
\end{figure}

In practice, ones always have to pass a 3 dimensional (3D) array of shape \{batch size, time steps, number of features\} as an input to the LSTM network. When output $y_t$ and all variables $x_{k,t}$ for $k = 1,2,3,...,K$ in the model specification are all sampled at the same frequency, both sequence stride $s$ (period between successive output sequences) and sampling rate $r$ (period between successive individual timesteps within variable sequences) are usually set to $1$ so as to utilize all available information. Nonetheless, when variables $x_{k,t}$ are all sampled at the same frequency mismatch ratio $m > 1$ times faster than an output $y_t$ for all $k=1,2,3,...,K$, such preprocessing strategy of $s=r=1$ will not work as the number of higher-frequency series $x_{k,t}$ is now $m$ times as many as the number of low-frequency series $y_t$. We denote this case the \textit{single frequency mismatch}. Subscript $\tau_k$ is used hereinafter to indicate possibly different sampling frequency of higher-frequency variables, and, consequently, the frequency mismatch ratio of individual higher-frequency series is indicated as $m_k$ for $k=1,2,3,...,K$. Fig.\ref{fig:fig1a} presents a data schedule for the case that a low-frequency output $y_t$ is sampled, say, quarterly and the frequency mismatch ratio between low-frequency output $y_t$ and each higher-frequency variable $x_{k,\tau_k}$ is 3 for all $k=1,2$; all higher-frequency variables $x_{k,\tau_k}$ are therefore sampled monthly and we require that the number of observations of higher-frequency $x_k$ is as exactly 3 times as the number of observations of low-frequency output $y_t$. A simple solution to apply LSTM with a model specification having only one frequency mismatch is then to preprocess the raw time series dataset with a sequence stride $s = 1$ and a sampling rate $r = m$ in order to match the occurrence of the higher-frequency variables $x_{k,t}$ to their low-frequency counterparts. Fig.\ref{fig:fig1b} show a simplified LSTM architecture with 5 timesteps in which the raw time series data are preprocessed with a sequence stride $s=1$, and a sampling rate $r=3$.

Unfortunately, the strategy of preprocessing time series data with sampling rate $r=m$ is not always workable when there are more than one frequency mismatch ratio in the model specification. Fig.\ref{fig:fig2a} shows an example of a quarterly series $y_t$, a monthly series $x_{\tau_1}^{1}$, and a bi-weekly series $x_{\tau_2}^{2}$. The frequency mismatch ratios are, therefore, 3 and 6 for the respective higher-frequency variables $x_1$ and $x_2$. As the data schedule shows, it is not possible to directly accommodate two higher-frequency variables with different frequency mismatch ratios into the same LSTM architecture. We denote this case the \textit{multiple frequency mismatches}. To deal with multiple frequency mismatches, we adopt the technique widely used in econometrics with mixed frequency time series data called the \textit{MIxed DAta Sampling} (MIDAS) approach \citep{ghysels2006predicting,ghysels2007midas}.

\begin{figure}[t]
    \centering
    \begin{subfigure}[b]{0.22\linewidth}
    \centering
    \tiny
    \begin{tikzpicture}[cell/.style={rectangle},
    row 1/.style={nodes={draw=none,fill=none}},
    space/.style={minimum height=2em,matrix of nodes,row sep=-\pgflinewidth,column sep=-\pgflinewidth, column 1/.style={nodes={anchor=west,minimum width=3em}}},
    B/.style = {decorate,decoration={brace,amplitude=5pt,pre=moveto,pre length=1pt,post=moveto,post length=1pt,raise=1mm}},
    text depth=0.5ex,nodes in empty cells]
    \matrix (second) [
    space,
    column 2/.style={anchor=west,nodes={cell,minimum width=3em}},
    column 3/.style={anchor=west,nodes={cell,minimum width=3em}},
    ]{
    $y_{t}$   & $x_{\tau_1}^{1}$    & $x_{\tau_2}^{2}$ \\
              &                     & $x_{1}^{2}$ \\
              & $x_{1}^{1}$         & $x_{2}^{2}$ \\
              &                     & $x_{3}^{2}$ \\
              & $x_{2}^{1}$         & $x_{4}^{2}$ \\
              &                     & $x_{5}^{2}$ \\
    $y_{1}$   & $x_{3}^{1}$         & $x_{6}^{2}$ \\
              &                     & $x_{7}^{2}$ \\
              & $x_{4}^{1}$         & $x_{8}^{2}$ \\
              &                     & $x_{9}^{2}$ \\
              & $x_{5}^{1}$         & $x_{10}^{2}$ \\
              &                     & $x_{11}^{2}$ \\
    $y_{2}$   & $x_{6}^{1}$         & $x_{12}^{2}$ \\
              &                     & $x_{13}^{2}$ \\
              & $x_{7}^{1}$         & $x_{14}^{2}$ \\
              &                     & $x_{15}^{2}$ \\
              & $x_{8}^{1}$         & $x_{16}^{2}$ \\
              &                     & $x_{17}^{2}$ \\
    $y_{3}$   & $x_{9}^{1}$         & $x_{18}^{2}$ \\
    \vdots    & \vdots              & \vdots \\
    $y_{n}$   & $x_{3n}^{1}$        & $x_{6n}^{2}$ \\
    \\
    \\
    };
    \draw[B] (second-1-2.north west) -- node[above=3mm] {Predictors} (second-1-3.north east);
    \draw[B] (second-1-1.north west) -- node[above=3mm] {Target} (second-1-1.north east);
    \end{tikzpicture}
    \subcaption{Raw data schedule}
    \label{fig:fig2a}
    \end{subfigure}
    \hfill
    \begin{subfigure}[b]{0.75\linewidth}
    \centering
    \tikzstyle{input}=[rectangle,draw,fill=pear!50,text width=13em,minimum height=2em]
    \tikzstyle{lstm}=[rectangle,draw,fill=pastelyellow,text width=5em,minimum height=2em]
    \tikzstyle{output}=[circle,draw,fill=awesome!50,text width=1em,minimum height=2em]
    \tikzstyle{truey}=[circle,draw,fill=cinnamon!50,text width=1em,minimum height=2em]
    \tikzstyle{arrow} = [thick,->,>=stealth]
    \tiny
    \begin{tikzpicture}[cell/.style={rectangle},
    row 1/.style={nodes={draw=none,fill=none}},
    space/.style={minimum height=2em,matrix of nodes,row sep=-\pgflinewidth,column sep=-\pgflinewidth, column 1/.style={nodes={anchor=west,minimum width=3em}}},
    B/.style = {decorate,decoration={brace,amplitude=5pt,pre=moveto,pre length=1pt,post=moveto,post length=1pt,raise=1mm}},
    text depth=0.5ex,nodes in empty cells]
    \matrix (second) [
    space,
    column 2/.style={anchor=west,nodes={cell,minimum width=2.8em}},
    column 3/.style={anchor=west,nodes={cell,minimum width=2.8em}},
    column 4/.style={anchor=west,nodes={cell,minimum width=2.8em}},
    column 5/.style={anchor=west,nodes={cell,minimum width=2.8em}},
    column 6/.style={anchor=west,nodes={cell,minimum width=2.8em}},
    column 7/.style={anchor=west,nodes={cell,minimum width=2.8em}},
    column 8/.style={anchor=west,nodes={cell,minimum width=2.8em}},
    column 9/.style={anchor=west,nodes={cell,minimum width=2.8em}},
    column 10/.style={anchor=west,nodes={cell,minimum width=2.8em}},
    column 11/.style={anchor=west,nodes={cell,minimum width=2.8em}},
    ]{
    $y_{t}$   & $x_{0/3}^{1}$ & $x_{1/3}^{1}$ & $x_{2/3}^{1}$ & $x_{3/3}^{1}$ & $x_{4/3}^{1}$ & $x_{0/6}^{2}$ & $x_{1/6}^{2}$ & $x_{2/6}^{2}$ & $x_{3/6}^{2}$ & $x_{4/6}^{2}$\\
    $y_{1}$ & $na$ & $na$ & $na$ & $na$ & $na$ & $x_{6}^{2}$ & $x_{5}^{2}$ & $x_{4}^{2}$ & $x_{3}^{2}$ & $x_{2}^{2}$ \\
    $y_{2}$ & $x_{6}^{1}$ & $x_{5}^{1}$ & $x_{4}^{1}$ & $x_{3}^{1}$ & $x_{2}^{1}$ & $x_{12}^{2}$ & $x_{11}^{2}$ & $x_{10}^{2}$ & $x_{9}^{2}$ & $x_{8}^{2}$ \\
    $y_{3}$ & $x_{9}^{1}$ & $x_{8}^{1}$ & $x_{7}^{1}$ & $x_{6}^{1}$ & $x_{5}^{1}$ & $x_{18}^{2}$ & $x_{17}^{2}$ & $x_{16}^{2}$ & $x_{15}^{2}$ & $x_{14}^{2}$ \\
    \vdots & \vdots & \vdots & \vdots & \vdots & \vdots & \vdots & \vdots & \vdots & \vdots & \vdots \\
    $y_n$ & $x_{3n}^{1}$ & $x_{3n-1}^{1}$ & $x_{3n-2}^{1}$ & $x_{3n-3}^{1}$ & $x_{3n-4}^{1}$ & $x_{6n}^{2}$ & $x_{6n-1}^{2}$ & $x_{6n-2}^{2}$ & $x_{6n-3}^{2}$ & $x_{6n-4}^{2}$ \\
    };
    \draw[B] (second-1-2.north west) -- node[above=3mm] {Frequency-aligned Predictors} (second-1-11.north east);
    \draw[B] (second-1-1.north west) -- node[above=3mm] {Target} (second-1-1.north east);
    \end{tikzpicture}
    \subcaption{Data schedule after frequency alignment}
    \label{fig:fig2b}
    \par\bigskip
    \begin{tikzpicture}[text centered,node distance=1cm and 0.25cm,auto]
    
    \node [output] (output2) {$\hat{y}_3$};
    \node [lstm,below of=output2] (cell52) {LSTM cell};
    \node [input,below of= cell52] (input62) {
    $x_{5}^{1},...,x_{9}^{1}$ , $x_{14}^{2},...,x_{18}^{2}$
    };
    \node [input,left of= input62,xshift=-2.5cm] (input52) {
    $x_{2}^{1},...,x_{6}^{1}$ , $x_{8}^{2},...,x_{12}^{2}$
    };
    \node [lstm,above of= input52] (cell42) {LSTM cell};
    \node [draw=none,right of= output2,xshift=-0.25cm] (o2) {$t=3$};
    \draw [arrow] (input62) -- (cell52);
    \draw [arrow] (input52) -- (cell42);
    \draw [arrow] (cell42) -- (cell52);
    \draw [arrow] (cell52) -- (output2);
    
    \end{tikzpicture}
    \subcaption{A simplified LSTM with 2 timesteps and the same lag structure}
    \label{fig:fig2c}
    \par\bigskip
    \begin{tikzpicture}[text centered,node distance=1cm and 0.1cm,auto]
    \node [output] (output2) {$\hat{y}_3$};
    \node [lstm,below of=output2] (cell52) {LSTM cell};
    \node [input,below of= cell52] (input62) {
    $x_{5}^{1},...,x_{9}^{1}$ , $x_{15}^{2},x_{16}^{2},x_{17}^{2}$
    };
    \node [input,left of= input62,xshift=-2.5cm] (input52) {
    $x_{2}^{1},...,x_{6}^{1}$ , $x_{9}^{2},x_{10}^{2},x_{11}^{2}$
    };
    \node [lstm,above of= input52] (cell42) {LSTM cell};
    \node [draw=none,right of= output2,xshift=-0.25cm] (o2) {$t=3$};
    \draw [arrow] (input62) -- (cell52);
    \draw [arrow] (input52) -- (cell42);
    \draw [arrow] (cell42) -- (cell52);
    \draw [arrow] (cell52) -- (output2);
    \end{tikzpicture}
    \subcaption{A simplified LSTM with 2 timesteps and different lag structures}
    \label{fig:fig2d}
    \end{subfigure}
\caption{Mixed frequency time series with multiple frequency mismatch ratios. (a) presents a raw data schedule in which the raw time series data are aligned to match with their actual period of occurrence. (b) presents a new data schedule in which the raw time series data are frequency-aligned following MIDAS approach. (c) presents the first feasible sequence of LSTM when timesteps are set to 2, and both variables take the same lag structure $j=\{0,1,2,3,4\}$. (d) also presents the first feasible sequence of LSTM when timesteps are set to 2, but a higher-frequency variable $x^1$ takes lag structure $j=\{0,1,2,3,4\}$ whereas a higher-frequency variables $x^2$ takes lag structure $j=\{1,2,3\}$.}
\end{figure}


\subsection{MIDAS approach}\label{sub:multiple}
To briefly introduce MIDAS regressions and simplify the exposition, we confine the discussion to a single variable $x_t$. Suppose that a higher-frequency series $x_t^{(m)}$ is sampled at $m$ times faster than a low-frequency series $y_t$. Hence, say, with a quarterly series $y_t$ and the frequency mismatch ratio $m=3$, a higher-frequency series $x_t$ is sampled monthly. Following the notation commonly used in \cite{ghysels2004midas,clements2008macroeconomic}, a simple linear MIDAS regression for a single explanatory variable and $h$-step-ahead forecasting can be written as:
\begin{equation}
\label{eqn:main}
    y_t = \alpha + \beta_{1}B(L^{1/m})x_{t-h}^{(m)} + \epsilon^{m}
\end{equation}
where $B(L^{1/m}) = \sum_{j=0}^{J}\omega_j L^{j/m}$ is a polynomial of length $J$ in the $L^{1/m}$ operator such that $L^{j/m}x_{t-h}^{(m)} = x_{t-h-j/m}^{(m)}$. In other words, the $L^{j/m}$ operator produces the value of $x_t$ lagged by $j/m$ periods (for simplicity, suppose we need to predict the end-of-quarter horizon, or $h=0$). Given the quarterly/monthly example, the above equation represents a projection of a quarterly time series $y_t$ onto a monthly time series $x_t^{m}$ up to $J$ monthly lags back. However, should a large number of lags of the $x_{t}^{(m)}$ be involved in a suitable polynomial $B(L^{1/m})$, ones would thus need to estimate many parameters. To avoid such parameter proliferation, Ghysels and his coauthors impose some restrictions upon the weight $\omega_j$ in the polynomial $B(L^{1/m})$. Following the pioneering work of \cite{ghysels2005there,ghysels2006predicting}, the most commonly used parameterization of $\omega_j$ is the second-order exponential Almon lag, or $\theta = (\theta_1,\theta_2)'$:
\begin{equation}
\label{eqn:weight}
    \omega_j(\theta) = \frac{exp(\theta_{1}j+\theta_{j}j^2)}{\sum_{j=1}^{J} exp(\theta_{1}j + \theta_{2}j^2)}
\end{equation}
where the vector of hyperparameters $\theta$ is unknown, $\omega_j(\theta) \in [0,1]$, and $\sum_{j=1}^{J} \omega_j(\theta)=1$. Practically, with this specification, a simple linear MIDAS regression model in (\ref{eqn:main}) can thus be written as:
\begin{equation}
\label{eqn:reduced}
    y_t = \alpha + \sum_{j=0}^{J} b_{j} x_{t-h}^{(m)} + \epsilon^{m}
\end{equation}
where $b_j = \beta_1 \omega_{j}(\theta)$ following \cite{andreou2013should} and \cite{ghysels2016mixed}.\footnote{This coefficient $b_j$ is also termed as \textit{MIDAS coefficient} when estimated using R package \textit{midasr} \citep{ghysels2016mixed}.} 
While model (\ref{eqn:reduced}) is a linear model in terms of variables, the nonlinear constraints imposed upon the weight $\omega_j$ result in nonlinearities with respect to the vector of hyperparameters $\theta$, and the model (\ref{eqn:reduced}) is estimated by nonlinear least squares (NLS) estimator.

The MIDAS specification is particularly useful for high mismatch applications in  financial time series data as shown by \cite{ghysels2005there,ghysels2006predicting}. However, for quarterly/monthly setting of macroeconomic applications in which $m$ is small and only a few higher-frequency lags are possibly needed to capture the weights, \cite{foroni2015unrestricted} show that the U-MIDAS model might be considered to be a good approximation to the true dynamic linear model.\footnote{Nonetheless, as also suggested by \cite{foroni2015unrestricted} that both MIDAS and U-MIDAS should be regarded as approximations to dynamic linear models, we should not expect one of these approaches to dominate with empirical data.} When estimating, the U-MIDAS also applies the same model (\ref{eqn:reduced}) as its MIDAS counterpart. However, without restriction Equation (\ref{eqn:weight}) being imposed, the U-MIDAS do not require NLS, and can be estimated by simple ordinary least square (OLS).

\subsection{Frequency alignment as a feature engineering process}
Practically, when estimating either MIDAS or U-MIDAS regression model, ones need to manipulate all variables observed at various frequencies and different length of observations. The process of transforming the higher-frequency variable $x_t$ into a low-frequency vector $(x_{3t},...x_{3t-2})'$ is denoted the \textit{frequency alignment} following \cite{ghysels2016mixed}. Suppose we assume that the monthly data in the current quarter has explanatory power, say, $J=2$ (i.e. a contemporaneous and two lags). This means that we need to model $y_t$ as a linear combination of variables $x_t,x_{t-{1/3}},x_{t-{2/3}}$ observed in each quarter $t$. Hence, the model (\ref{eqn:main}) can be written in matrix notation as:

\begin{gather}\label{eqn:falign}
\begin{bmatrix}
y_{1} \\ \vdots \\ y_{n} 
\end{bmatrix}
=
\alpha +
\begin{bmatrix}
x_{3}  & x_{2} & x_{1} \\
\vdots & \ddots & \vdots \\
x_{3n} & x_{3n-1} & x_{3n-2}
\end{bmatrix}
\begin{bmatrix}
b_{0} \\ \vdots \\ b_{2}
\end{bmatrix}
+
\begin{bmatrix}
\epsilon_{1} \\ \vdots \\ \epsilon_{n}
\end{bmatrix}
\end{gather}

Since all the variables are now observed at the same frequency, this process enables a MIDAS regression to directly accommodate variables sampled at different frequencies and turns it into a classical time series regression. Although the Equation (\ref{eqn:falign}) presents a simple case of a single frequency mismatch, the procedure is very generalizable. Fig.\ref{fig:fig2a} presents a case of quarterly output $y_t$ and two frequency mismatches $m_1 = 3$ (monthly) and $m_2 = 6$ (bi-weekly). Suppose we assume that a contemporaneous and four lags (i.e. $j=0,1,2,3,4$) of both variables have explanatory power. The resulting data schedule after preprocessing raw time series with frequency alignment process has now additional eight features as shown in Fig.\ref{fig:fig2b}. Despite the fact that $x_{3}^{1},x_{2}^{1},x_{1}^{1}$ exist, their corresponding entries of the variable $x_{\tau}^{1}$ in the row $t=1$ are intentionally shown as non-available because a lag order of 4 is beyond its sampling frequency $m_{1}=3$ which causes the frequency alignment at $t=1$ to fail. The guiding model specification can be written in matrix notation as:

\begin{equation}
\begin{aligned}
\begin{bmatrix}
y_{2} \\ \vdots \\ y_{n} 
\end{bmatrix}
=
\alpha &+
\begin{bmatrix}
x_{6}^{1}  & x_{5}^{1} & x_{4}^{1} & x_{3}^{1} & x_{2}^{1} \\
\vdots & \vdots & \vdots & \vdots & \vdots \\
x_{3n}^{1} & x_{3n-1}^{1} & x_{3n-2}^{1} & x_{3n-3}^{1} & x_{3n-4}^{1}
\end{bmatrix}
\begin{bmatrix}
b_{0}^{1} \\ \vdots \\ b_{4}^{1}
\end{bmatrix} \\
&+
\begin{bmatrix}
x_{6}^{2}  & x_{5}^{2} & x_{4}^{2} & x_{3}^{2} & x_{2}^{2} \\
\vdots & \vdots & \vdots & \vdots & \vdots \\
x_{3n}^{2} & x_{3n-1}^{2} & x_{3n-2}^{2} & x_{3n-3}^{2} & x_{3n-4}^{2}
\end{bmatrix}
\begin{bmatrix}
b_{0}^{2} \\ \vdots \\ b_{4}^{2}
\end{bmatrix}
+
\begin{bmatrix}
\epsilon_{2} \\ \vdots \\ \epsilon_{n}
\end{bmatrix}
\end{aligned}
\end{equation}

The frequency alignment process can also be generalized to the case of multiple frequency mismatches with different length of lag vector. Moreover, the vector of lags included for each variable can also differently start from a given order rather than from a contemporaneous lag ($j=0$). Fig.\ref{fig:fig2c} and \ref{fig:fig2d} present the first feasible sequence of LSTM with 2 timesteps in which the raw time series data are preprocessed with frequency alignment. The former presents the case when both higher-frequency variables take the same lag structure (contemporaneous with 4 lags) as shown in Fig.\ref{fig:fig2b}. The latter provides an example of different lag structures where a higher-frequency variable $x^1$ takes lag structure $j=\{0,1,2,3,4\}$ and a higher-frequency variables $x^2$ takes lag structure $j=\{1,2,3\}$.

Setting aside the estimation and restrictions, both MIDAS and U-MIDAS approaches rely on the frequency alignment preprocessing process and explicitly treat lagged observations (at particular orders) as their variables. Put it differently, from a machine learning perspective, the frequency alignment can simply be viewed as an alternative feature engineering process. Hence, ones can directly adopt the U-MIDAS scheme into the LSTM, and estimate all parameters as its usual using standard library such as Keras in Python.


\section{Monte Carlo Simulations}
In this section, we assess via Monte Carlo simulations the predictive performance to illustrate the potentials of adopting the unrestricted lag polynomials projection in LSTM architectures. Specifically, we benchmark the out-of-sample nowcasting (i.e. 1-, 2-, and 3-month ahead) and forecasting (i.e. 6-, 9-, and 12-month ahead) performance of LSTM against the MIDAS and U-MIDAS regression models.

\subsection{Experimental design}
The simulation exercise is designed in a way to discuss its application in macroeconomic forecasting, in particular forecasting quarterly GDP growth. To be relevant for empirical application in the next section, we focus on the case of quarterly/monthly ($m=3$) data with small sample size environment. We follow the simulation design in \cite{foroni2015unrestricted} that directly uses the restricted MIDAS regression as the data generating process (DGP). The guiding DGP is explicitly defined as: \begin{equation}
\begin{gathered}
y_{t+h} = \alpha + \sum_{k=1}^{K} \sum_{j=0}^{J} b_{j}^{k} x_{k,t-j}^{(3)} + \epsilon_{t+h}
\\
\epsilon_{t+h} \sim N(0,1), \forall t = 1,...,T
\end{gathered}
\end{equation}
For simplicity, a contemporaneous and the maximum lag order $J=11$ lags (i.e. $j=0,1,2,...,11$) are used for all higher-frequency variables. The parameterized lag coefficients $b_{j}^{k}(j;\theta) = \beta_{j}^{k} \omega_{kj}$, where the weights, $\omega_{ij}$, are generated by the normalized exponential Almon lag polynomial (see \cite{ghysels2016mixed} for more detail) and defined as:
\begin{equation}
\omega_{kj} = \frac{exp(\theta_1^{k}j+\theta_{2}^{k}j^{2})}{\sum_{j=1}^{J} exp(\theta_1^{k}j+\theta_{2}^{k} j^{2})}
\end{equation}
The monthly variables are generated as either AR(1) process with persistence equal to 0.9 or $x_{k,\tau} \sim N(0,1)$. The sample size (expressed in the low-frequency unit) is $T = \{50,80\}$. Four different DGPs of low-frequency output $y_t$ are therefore simulated and evaluated. For the sake of comparison, all experiments are designed to be the single frequency mismatch $m=3$ class. All simulations take the number of higher-frequency variables $K = 3$. The parameter pairs of $\theta = (\theta_1,\theta_2)'$ are from the sets $\theta_1 = 0.7$ and $\theta_2 = \{-0.025,-0.1,-0.5\}$ following parameters commonly used in the MIDAS literature (e.g. \cite{ghysels2016mixed,foroni2015unrestricted}). To some extent, the simulations are in favor of the MIDAS as a small number of the most relevant variables are practically included in the MIDAS regression model so as to preserve parsimony. On the contrary, a large vector of lags will directly affect the forecasting performance of the U-MIDAS regression model as the number of parameters to be estimated is large with respect to the sample size.

We assess the forecasting performance of the proposed data preprocessing strategies discussed in previous section. For each DGP, all three higher-frequency variables generated are taken as the input data. The input data of the first two LSTM models are preprocessed using the sampling rate $r = m$ strategy as discussed in section \ref{sub:single}. The timesteps are set to 6 and 12, respectively. We denote these models \textit{Sampling-aligned} LSTM (SA-LSTM). To apply the U-MIDAS scheme, the input data of the other three models are preprocessed using MIDAS frequency alignment strategy as discussed in section \ref{sub:multiple}. We denote these models \textit{Frequency-aligned} LSTM (FA-LSTM). The FA-LSTM models are set to have the lag structure starting from the lag order zero (i.e. a contemporaneous lag) and take the vector of lag orders of [0:2], [0:5], and [0:11], respectively. Since they take all higher-frequency variables ($K = 3$), the number of variables used for each FA-LSTM model are $3 \times (J+1) = 9, 18,$ and $36$, respectively. The timesteps of 4, 2, and 1 are applied for three FA-LSTM models.

We conduct a grid search to select the best hyperparameter combination of epochs, dropout, batch size, and the number of LSTM memory cells for each LSTM model in each forecasting horizon. As the objective of the simulation is not to search for the highest forecasting accuracy, only the small set of epochs = $\{25,50\}$, dropout = $\{0,0.4\}$, and batch size = $\{1,\lceil{input/10}\rceil,\lceil{input/2}\rceil\}$, where $input$ is the effective size of the training data and $\lceil\cdot\rceil$ denotes the ceiling function, are examined. For the number of LSTM memory cells, a set of $\{16,32,64,128\}$ is examined for FA-LSTM models, and a set of $\{8,16,32\}$ is examined for SA-LSTM models. We simply exploit the first twenty simulation replications (i.e. $R = \{1,2,3,...,20\}$) that already prepared for the experiment as the dataset. For each simulation replication, the ratio of 60:40 is applied to split the data into training and validation samples. To minimize computational time, we evaluate the predictive performance of each hyperparameter combination on the validation samples using a fixed forecast scheme. The average root mean squared forecast error (RMSFE) over these twenty simulation replications are then compared across all hyperparameter combinations. Table \ref{table:mc_para} presents hyperparameter combinations that achieves the lowest average RMSFE for all forecasting horizons and DGPs.

We benchmark the out-of-sample forecasting performance of the LSTM models against two alternatives: MIDAS and U-MIDAS regressions. We consider the model specifications that are parsimonious but possibly misspecified functional forms. Thus, only two higher-frequency variables are used in the model specification. For a given forecasting horizon, the same model specification is applied for both MIDAS and U-MIDAS models. All (U-)MIDAS regression models are estimated by midasr package \citep{ghysels2016mixed} in R. The best model specification for each forecasting horizon is thoroughly searched over the set of potential models defined by all combinations of the number of lags (2 to 12 lags), higher-frequency variable pair (3 possible pairs), and weighting functions (normalized and non-normalized exponential Almon lag polynomials). For simplicity, the lowest lags of high-frequency variables are simply set to be equal to the respective higher-frequency forecasting horizon $h_m$. An automatic selection is performed using the flexible function available in the midasr package. As applied in the LSTM models, the ratio of 60:40 is applied to split the data into training and testing data. The fixed model option is also set for forecasting. We select the model that achieves the lowest mean squared forecast error (MSE). 

The out-of-sample forecasting performance is assessed across a series of leading index $d = \{m-1,m-2,0,0,0,0\}$, where $d$ represents the number of higher-frequency observations that are available before the release of low-frequency output. Given the fixed frequency mismatch $m=3$ in the experiments, this corresponds to higher-frequency forecasting horizons $h_m = \{1,2,3,6,9,12\}$ and a sequence of low-frequency forecasting horizons $h = \{1,1,1,2,3,4\}$. For each simulation replication, the ratio of 60:40 is also applied to split the data into training observations $T_1$ and evaluation observations $T_2$. Given the sample size $T=\{50,80\}$, the evaluation observations $T_2 = \{20,32\}$ are held out and used as out-of-sample data, and the rest $T_1 = T - T_2$ observations as training data. To evaluate forecasting performance, we assume that the information up to period $T_{1} + \Omega$, where $\Omega = 0,1,2,...,T_{2}-h$, is available for low-frequency output and $T_{1} + \Omega + d/m$ for higher-frequency variables. We then conduct one-horizon-ahead rolling forecast without re-estimation. This yields a forecast of the low-frequency output $y_t$ for the $h_m$ higher-frequency periods ahead for the out-of-sample observations $\hat{y}_{T_1+\Omega+h_m}$. The corresponding squared forecast error $SE = (y_{T_1+\Omega+h_m} - \hat{y}_{T_1+\Omega+h_m})^2$ is thus used to calculate the RMSFE over the evaluation sample for each simulation replication. The number of simulations are fixed at $R=1000$. Since the random initial conditions in the LSTM architecture may result in very different results, we estimate all LSTM models twice for each replication. The out-of-sample RMSFEs reported in the next section are therefore calculated by averaging over $2 \times 1000$ replications. All LSTM models are estimated by Keras library in Python. 

\subsection{Simulation results}

Table \ref{table:monteNor}-\ref{table:monteAR} report detailed results of Monte Carlo simulation experiment. The U-MIDAS in simulation scenarios performs unsurprisingly poorly as the number of parameters to be estimated is large with respect to the sample size. As such, the restricted MIDAS substantially outperforms its unrestricted counterpart across all forecasting horizons and sample sizes. We apply the \cite{diebold1995comparing} test statistics (hereinafter, DM test) to examine whether the difference of RMSFE from the two competing models are statistically significant. Since all U-MIDAS perform poorly, we only assess the statistical difference of forecast accuracy between the restricted MIDAS and the LSTM for each forecasting horizon.

\begin{table}[b!]
\centering
\begin{threeparttable}
\caption{Out-of-sample RMSFE results for Monte Carlo simulation (DGP: $x_{k,\tau} \sim N(0,1)$)}
\label{table:monteNor}
\begin{tabular}{@{}cccclcclccc@{}}
\toprule
\multirow{2}{*}{\begin{tabular}[c]{@{}c@{}}Raw Obs.\end{tabular}} & \multirow{2}{*}{$h_m$} & \multirow{2}{*}{MIDAS} & \multirow{2}{*}{U-MIDAS} &  & \multicolumn{2}{c}{SA-LSTM} &  & \multicolumn{3}{c}{FA-LSTM} \\ \cmidrule(lr){6-7} \cmidrule(l){9-11} &  &  &  &  & \multicolumn{1}{l}{{[}6, 0:0{]}} & \multicolumn{1}{l}{{[}12, 0:0{]}} &  & \multicolumn{1}{l}{{[}4, 0:2{]}} & \multicolumn{1}{l}{{[}2, 0:5{]}} & \multicolumn{1}{l}{{[}1, 0:11{]}} \\ \midrule
50 & 1 & 1.165 & 1.446 &  & \textbf{1.095}* & \textbf{1.101} &  & \textbf{1.119} & \textbf{1.125} & \textbf{1.138} \\[1ex]
   &   & (0.201) & (0.299) &  & (0.171) & (0.176) &  & (0.179) & (0.178) & (0.180) \\[1ex]
   & 2 & 1.170 & 1.387 &  & \textbf{1.104}* & \textbf{1.121} &  & \textbf{1.124} & \textbf{1.129} & \textbf{1.136} \\[1ex]
   &   & (0.200) & (0.267) &  & (0.173) & (0.195) &  & (0.178) & (0.182) & (0.181)  \\[1ex]
   & 3 & 1.171 & 1.359 &  & \textbf{1.114}* & \textbf{1.129} &  & \textbf{1.127} & \textbf{1.132} & \textbf{1.148} \\[1ex]
   &   & (0.195) & (0.260) &  & (0.176) & (0.187) &  & (0.180) & (0.180) & (0.183)  \\[1ex]
   & 6 & 1.176 & 1.319 &  & \textbf{1.121}* & \textbf{1.130} &  & \textbf{1.130} & \textbf{1.137} & \textbf{1.147} \\[1ex]
   &   & (0.199) & (0.250) &  & (0.175) & (0.186) &  & (0.181) & (0.183) & (0.186)  \\[1ex]
   & 9 & 1.175 & 1.308 &  & \textbf{1.112}* & \textbf{1.125} &  & \textbf{1.129} & \textbf{1.142} & \textbf{1.150} \\[1ex]
   &   & (0.248) & (0.243) &  & (0.175) & (0.190) &  & (0.180) & (0.185) & (0.185)  \\[1ex]
   &12 & 1.217 & 2.571 &  & \textbf{1.114}* & 1.207 &  & \textbf{1.135} & \textbf{1.146} & \textbf{1.155} \\[1ex]
   &   & (0.220) & (0.997) &  & (0.175) & (0.262) &  & (0.181) & (0.184) & (0.183)  \\[1ex] \midrule
80 & 1 & 1.120   & 1.167  &  & \textbf{1.158} & \textbf{1.146} &  & \textbf{1.106}* & 1.117 & \textbf{1.130} \\[1ex]
   &   & (0.145) & (0.155)&  & (0.175) & (0.189) &  & (0.140) & (0.143) & (0.146) \\[1ex]
   & 2 & 1.134 & 1.180 &  & \textbf{1.116} & \textbf{1.104}* &  & \textbf{1.117} & \textbf{1.126} & 1.134 \\[1ex]
   &   & (0.145) & (0.156)&  & (0.143) & (0.140) &  & (0.141) & (0.147) & (0.147) \\[1ex]
   & 3 & 1.136 & 1.216 &  & \textbf{1.106}* & \textbf{1.111} &  & \textbf{1.122} & \textbf{1.128} & \textbf{1.132} \\[1ex]
   &   & (0.151) & (0.169)&  & (0.140) & (0.142) &  & (0.143) & (0.146) & (0.146) \\[1ex]
   & 6 & 1.146 & 1.261 &  & \textbf{1.111}* & \textbf{1.168} &  & \textbf{1.126} & \textbf{1.130} & \textbf{1.139} \\[1ex]
   &   & (0.153) & (0.180)&  & (0.144) & (0.183) &  & (0.147) & (0.147) & (0.146) \\[1ex]
   & 9 & 1.155 & 1.322 &  & \textbf{1.117} & \textbf{1.114}* &  & \textbf{1.125} & \textbf{1.136} & \textbf{1.142} \\[1ex]
   &   & (0.154) & (0.200)&  & (0.143) & (0.142) &  & (0.145) & (0.146) & (0.150) \\[1ex]
   & 12& 1.153 & 1.389 &  & \textbf{1.116}*& \textbf{1.122}  &  & \textbf{1.135} & \textbf{1.138} & \textbf{1.147}  \\[1ex]
   &   & (0.155) & (0.214)&  & (0.142) & (0.146) &  & (0.147) & (0.148) & (0.146)\\[1ex] \bottomrule
\end{tabular}
\begin{tablenotes}[flushleft]
      \small
      \item \textit{Notes}: This table reports out-of-sample RMSFE results for simulation exercises. Except the first two columns, the table reports average RMSFE across 1,000 replications $\times$ 2 estimations per replication. Standard deviation is reported in parentheses. * indicates the lowest average RMSFE in the given forecasting horizon. Bold text indicates that the null hypothesis of the two methods having the same forecast accuracy is rejected at the 5\% significance level using \cite{diebold1995comparing} test statistics. The timesteps and lag specification of the input data, which are fixed across all forecasting horizons, are shown for each LSTM model. Since the SA-LSTM models take no lagged variables, the lag specification of [0:0] is simply shown. The optimal batch size for each LSTM model in each forecasting horizon are separately reported in Table \ref{table:mc_para}. Given the number of higher-frequency variables $K = 3$, the total number of variables used in each LSTM model are $K \times J+1 = 3,3,9,18,36$, respectively.
      
\end{tablenotes}
\end{threeparttable}
\end{table}

\begin{table}[b!]
\centering
\begin{threeparttable}
\caption{Out-of-sample RMSFE results for Monte Carlo simulation (DGP: $x_{k,\tau} = 0.9 x_{k,\tau-1}+\epsilon_{\tau}$)}
\label{table:monteAR}
\begin{tabular}{@{}ccccccccccc@{}}
\toprule
\multirow{2}{*}{Raw obs.} & \multirow{2}{*}{$h_m$} & \multirow{2}{*}{MIDAS} & \multirow{2}{*}{U-MIDAS} &  & \multicolumn{2}{c}{SA-LSTM}  &  & \multicolumn{3}{c}{FA-LSTM} \\ \cmidrule(lr){6-7} \cmidrule(l){9-11}
    &  &  &  &  & {[}6, 0:0{]} & {[}12, 0:0{]} &  & {[}4, 0:2{]} & {[}2, 0:5{]} & {[}1, 0:11{]} \\ \midrule
50  & 1 & 2.066 & 2.311 &  & \textbf{2.108} & \textbf{2.633} &  & 1.922 & \textbf{1.634}* & \textbf{1.741} \\[1ex]
    &   & (0.616) & (0.696) &  & (1.395) & (3.581) &  & (0.924) & (0.559) & (0.538)\\[1ex]
    & 2 & 2.142   & 3.809   &  & \textbf{2.229} & \textbf{3.013}&  & \textbf{2.114} & \textbf{1.999} & \textbf{1.963}* \\[1ex]
    &   & (0.713) & (1.523) &  & (1.359) & (6.200) &  & (0.938) & (0.646) & (0.599) \\[1ex]
    & 3 & 2.251   & 3.440   &  & \textbf{2.014} & \textbf{3.128}&  & \textbf{2.193} & \textbf{1.989}* & \textbf{2.108} \\[1ex]
    &   & (0.729) & (1.190) &  & (0.738) & (6.857) &  & (0.750) & (0.625) & (0.659) \\[1ex]
    & 6 & 2.415   & 3.501   &  & \textbf{2.495} & \textbf{2.141}*&  & 2.469 & \textbf{2.403} & \textbf{2.507} \\[1ex]
    &   & (0.815) & (1.193) &  & (1.226) & (0.975) &  & (0.816) & (0.768) & (0.782) \\[1ex]
    & 9 & 2.787   & 3.630   &  & \textbf{2.616}*& \textbf{4.194}&  & \textbf{2.686} & \textbf{2.743} & 2.798 \\[1ex]
    &   & (1.044) & (1.250) &  & (0.849) & (9.771) &  & (0.894) & (0.905) & (0.934) \\[1ex]
    & 12& 2.826   & 3.215   &  & \textbf{2.750}*& \textbf{2.979}&  & \textbf{2.859} & \textbf{2.870} & \textbf{3.005} \\[1ex]
    &   & (0.938) & (0.995) &  & (0.900) & (3.350) &  & (0.993) & (0.978) & (1.066) \\[1ex] \midrule
80  & 1 & 1.901   & 2.079   &  & 1.780   & \textbf{2.663} &  & \textbf{1.530} & \textbf{1.524}* & \textbf{1.571} \\[1ex]
    &   & (0.449) & (0.48)  &  & (0.813) & (4.020) &  & (0.416) & (0.376) & (0.360) \\[1ex]
    & 2 & 1.947   & 2.094   &  & 1.842   & \textbf{2.970} &  & \textbf{1.750} & \textbf{1.602}* & \textbf{1.653} \\[1ex]
    &   & (0.447) & (0.469) &  & (0.847) & (4.437) &  & (0.474) & (0.365) & (0.384) \\[1ex]
    & 3 & 2.026   & 2.079   &  & \textbf{1.842} & \textbf{2.557} &  & \textbf{1.824} & \textbf{1.764}* & \textbf{1.830} \\[1ex]
    &   & (0.457) & (0.443) &  & (0.784) & (2.965) &  & (0.467) & (0.390) & (0.447) \\[1ex]
    & 6 & 2.156*  & 2.187   &  & 2.183   & \textbf{2.213} &  & \textbf{2.372} & \textbf{2.223} & \textbf{2.388} \\[1ex]
    &   & (0.462) & (0.470) &  & (0.536) & (1.466) &  & (0.707) & (0.536) & (0.594) \\[1ex]
    & 9 & 2.423*  & 2.656   &  & \textbf{2.547} & \textbf{3.460} &  & \textbf{2.570} & \textbf{2.579} & \textbf{2.631} \\[1ex]
    &   & (0.575) & (0.608) &  & (0.690) & (5.761) &  & (0.656) & (0.660) & (0.672) \\[1ex]
    & 12& 2.617   & 2.754   &  & 2.648   & 2.535*  &  & \textbf{2.762} & \textbf{2.737} & \textbf{2.794} \\[1ex]
    &   & (0.644) & (0.636) &  & (0.649) & (1.158) &  & (0.774) & (0.718) & (0.743) \\[1ex] \bottomrule
\end{tabular}
\begin{tablenotes}[flushleft]
      \small 
      \item \textit{Notes}: This table reports out-of-sample RMSFE results for simulation exercises. Except the first two columns, the table reports average RMSFE across 1,000 replications $\times$ 2 estimations per replication. Standard deviation is reported in parentheses. * indicates the lowest average RMSFE in the given forecasting horizon. Bold text indicates that the null hypothesis of the two methods having the same forecast accuracy is rejected at the 5\% significance level using \cite{diebold1995comparing} test statistics. The timesteps and lag specification of the input data, which are fixed across all forecasting horizons, are shown for each LSTM model. Since the SA-LSTM models take no lagged variables, the lag specification of [0:0] is simply shown. The optimal batch size for each LSTM model in each forecasting horizon are separately reported in Table \ref{table:mc_para}. Given the number of higher-frequency variables $K = 3$, the total number of variables used in each LSTM model are $K \times J+1 = 3,3,9,18,36$, respectively.
      
\end{tablenotes}
\end{threeparttable}
\end{table}

When DGP of higher-frequency variables is $x_{k,\tau} \sim N(0,1)$, the LSTM with either sampling alignment (SA-LSTM) or frequency alignment (FA-LSTM) data preprocessing strategy outperform the MIDAS model across all higher-frequency forecasting horizons $h_m = \{1,2,3,6,9,12\}$ and sample sizes $T=\{50,80\}$. Almost all of the DM tests are statistically significant at the level of 5\%. Although the SA-LSTM models performer best in most of the simulation scenarios generated by the normal distribution process, the forecasting performance of the FA-LSTM model does not substantially differ from that of the SA-LSTM model. However, these findings are not robust to the DGP of the AR(1) process with persistence equal to 0.9. All methods perform worse with persistent higher-frequency variables across all simulation scenarios. They all obtain larger RMSFE and larger standard deviation. The SA-LSTM model seems to be affected more compared to the other methods. Most of the forecasts produced by the SA-LSTM model perform significantly worse than the FA-LSTM model and the restricted MIDAS model. Although most of the SA-LSTM models still achieve the lowest average RMSFE, especially when $h_m = \{6,9,12\}$, many of them perform worse than the competing models both in terms of larger RMSFE and larger standard deviation. Interestingly, the FA-LSTM performs much better than the competing models when $h_m = \{1,2,3\}$. This simulation evidence encourages the use of LSTM with frequency alignment data preprocessing when ones need to nowcast the low-frequency series of interest.

To sum up, even in a set up favorable to the restricted MIDAS, the LSTM with either sampling alignment or frequency alignment data preprocessing still yields a better forecasting performance than MIDAS. Simulation results indicate that the sampling alignment strategy, which we simply adjust the sampling rate in the commonly used data preprocessing procedure, is an interesting strategy when the mixed frequency data problem has only one frequency mismatch. Nonetheless, the frequency alignment strategy provides more flexibility when ones need to forecast the low-frequency data series with multiple frequency mismatches. Although we cannot expect one of the approaches to dominate with empirical data, the simulation results seem to support the use of LSTM with U-MIDAS scheme when ones need to nowcast the macroeconomic low-frequency data series, such as GDP growth, in the quarterly/monthly setting.


\section{Empirical application to Thai GDP growth}

This section presents an empirical application of the proposed models to real world data. Specifically, we predict quarterly growth rate of Thai real GDP series using a vast array of macroeconomic indicators both quarterly and monthly, and assess the predictive performance of the SA-LSTM and FA-LSTM models across six higher-frequency forecasting horizons (i.e. $h_m=1,2,3,6,9,12$). The predictive performance of the proposed LSTM modes are benchmarked against variety of competing models. Our empirical exercises focus on non-seasonally adjusted QGDP series and mimics the real-time situation in the sense that the data publication delays are replicated; hence, this is a pseudo real-time forecast evaluation. To be policy relevant, we additionally recalculate the GDP growth as an annual rate using the quarterly forecasts computed by the FA-LSTM model. The annualized QGDP growth forecasts are then compared with the publicly available forecasts produced by the Bank of Thailand model at various forecasting horizons.

\subsection{Data}

Regarding the macroeconomic series, we first collect all quarterly and monthly series available in the \href{https://www.bot.or.th/English/Statistics/Pages/default.aspx}{BOT} database. The time of the first available observation differs from series to series, and all of them are available up to the end of 2020. The longest series start at 1993M1 and 1993Q1 for monthly and quarterly series, respectively. To prevent very short samples, we filter out the series of which the first period is available later than the cut-off period of 2001M1. The final dataset of variables contains five quarterly and 35 monthly series. We impute zero values for the series that start later than 1993M1 for the monthly series and 1993Q1 for the quarterly series. The full list of the series with further details appear in the Online Appendix B. The quarterly real GDP (QGDP) series in chained volumes from 1993Q1 to 2020Q4 are available from the \href{https://www.nesdc.go.th/nesdb_en/main.php?filename=national_account}{Office of the National Economic and Social Development Council} (NESDC). The growth rate is then calculated directly as $100 \times (QGDP_{t+1}-QGDP_{t})/QGDP_{t}$.

Except QGDP series, all macroeconomic series are final revised data and supposedly published with a delay of up to one month according to the observed typical pattern between the end of the reference period and the date of the respective release. This implies that if we need to forecast the QGDP growth for 2019Q1 three months before the quarter ends (i.e. $h_m=3$), we will need data up to the end of 2018M12. Given that all macroeconomic series are published with a delay, it effectively means that the contemporaneous lags are 2018M11 and 2018Q3 for monthly and quarterly series, respectively. The QGDP series is periodically published within eight weeks after the end of the reference period; hence, it is available for forecasting at the beginning of the third month of the next quarter. Again, for example, if we need to forecast the QGDP growth at the end of 2019Q1, the first available lag to be used as the autoregressive (AR) term of the QGDP growth series is 2018Q3 for $h_m=2,3$, and 2018Q4 for $h_m=1$.

\subsection{Empirical design}

We evaluate the predictive performance of the LSTM using U-MIDAS scheme (i.e. FA-LSTM) and the SA-LSTM models on the out-of-sample data, containing 52 quarters, from 2008Q1 to 2020Q4. Six projections are computed for six higher-frequency forecasting horizons ($h_m=1,2,3,6,9,12$) for each QGDP observation of the out-of-sample period. All competing models are recursively configured and estimated at the beginning of each quarter using the information up to the end of the last available month for a given forecasting horizon $h_m$. For the first forecast at the end of 2008Q1, for example, the competing models are configured and estimated at the beginning of 2008M3 ($h_m=1$), 2008M2 ($h_m=2$), 2008M1 ($h_m=3$), 2007M10 ($h_m=6$), 2007M7 ($h_m=9$), and 2007M4 ($h_m=12$) using the data up to the end of 2008M2, 2008M1, 2007M12, 2007M9, 2007M6, and 2007M3, respectively.

As a natural comparison, we compare out-of-sample forecasts of the proposed models with the univariate LSTM, denoted as UNI-LSTM, model. The next section shows that, before the COVID-19 pandemic, this relatively much simpler univariate LSTM benchmark turns out to be a very strong competitor to the proposed models. Following the MIDAS literature for short-term GDP growth predictions \citep{marcellino2010factor,babii2021machine}, the simple benchmark AR(1) model is also used as an additional benchmark for assessing the short-term forecasting ($h_m=1,2,3$) performance of the proposed models. Although the LSTM effectively capture the sequential information in the input data, the nature of the economy is highly dynamic and its structure may substantially change over time. If that is the case, the advantages of LSTM architecture having both a short-term and a long-term memory might be trivial. We therefore report the results for the artificial neural networks (ANN) model using the U-MIDAS scheme, denoted U-MIDAS-ANN, as another competing model.

To assess the predictive performance of the models, we compare RMSFE of the proposed models with the benchmark models in a recursive forecasting exercise. We additionally relate the mean squared forecast error (MSFE) of the proposed models to the variance of QGDP growth over the evaluation period following \cite{marcellino2010factor}. A ratio of less than one indicates that the predictive performance of a model is to some extent informative. We also apply the DM test statistics \citep{diebold1995comparing} to examine whether the difference of RMSFE from the two competing models are statistically significant.

\subsection{Model specification and hyperparameter tuning}

The following ADL-MIDAS-like equation is applied as a \textit{guiding} model specification for the proposed FA-LSTM and the alternative U-MIDAS-ANN models for all forecasting horizons:
\begin{equation}
\begin{gathered} \label{eq:guide_fa}
y_{t+h} = \alpha + \sum_{j=0}^{P_{Q}-1} \lambda_{j} y_{t-j} + \sum_{k=1}^{K_Q} \sum_{j=0}^{P_{Q}-1} b_{j}^{k} x_{k,t-j}^{Q} + \sum_{k=1}^{K_M} \sum_{j=0}^{P_{M}-1} b_{j}^{k} x_{k,t-j}^{M} + \epsilon_{t+h}
\end{gathered}
\end{equation}
where $x_{k}^{Q}$ and $x_{k}^{M}$ represents low-frequency quarterly and higher-frequency monthly variable, respectively. The guiding specification (\ref{eq:guide_fa}) involves a contemporaneous and $P_{Q}-1$ lags of $x_{t}^{Q}$ and a contemporaneous and $P_{M}-1$ lags of $x_{t}^{M}$. Following \cite{clements2008macroeconomic}, we add the AR dynamics $y_{t-j}$ to a guiding model specification so as to provide a more flexible dynamic specification. This approach is, from the econometric point of view, useful to handle eventual serial correlation in the idiosyncratic components \citep{marcellino2010factor}. For simplicity, we apply the same lag orders for all variables having the same frequency mismatch ratio; hence, quarterly variables and AR dynamics $y_t$ included in a given model are always forced to have the same contemporaneous and $P_{Q}-1$ lags. Equation (\ref{eq:guide_sa}) presents a guiding model specification for the SA-LSTM model:
\begin{equation}
\begin{gathered} \label{eq:guide_sa}
y_{t+h} = \alpha + \sum_{k=1}^{K_M} \sum_{j=0}^{P_{M}-1} b_{j}^{k} x_{k,t-j}^{M} + \epsilon_{t+h}
\end{gathered}
\end{equation}
It differs from that of the FA-LSTM model since the SA-LSTM model can not take more than one frequency mismatch ratio in the model specification. This guiding model specification is used for all nowcasting and forecasting horizons.

Since a choice of model specification (i.e. the number of lag orders $P$ and a set of $K$ variables) would consequently affect a selection of hyperparameters, both optimal number of lag orders and hyperparameters are simultaneously selected using the grid search method for all competing models, except the benchmark UNI-LSTM and AR(1) models. For each combination of lag orders and hyperparameters, we first use the Least Absolute Shrinkage and Selection Operator (LASSO) regression \citep{tibshirani1996regression} to select a set of variables which minimize 5-fold time series cross validation (i.e. the previous four quarters of a given out-of-sample forecast) in terms of MSFE. The resulting variables are presumably considered as the best available set of variables for forecasting the next out-of-sample period. Second, we apply the ratio of 80:20 to split the available information into training and validation samples, fit the model using training samples, and then compute the forecasts for the validation samples using a fixed forecast scheme. Lastly, we compare the predictive performance on validation samples of all combinations in terms of RMSFE. Since the random initial conditions in the algorithm may result in very different forecasts, each combination is estimated three times and calculate an average RMSFE. We select a combination of lag orders and hyperparameters which achieves the lowest RMSFE. Note that the whole model configuration process is recursively conducted in every out-of-sample forecast; hence, the combination of variables, maximal lag orders, and hyperparameters are allowed to change over time. Since the number of foreign tourists unprecedentedly drop to zero in April 2019 due to the travel restrictions imposed around the world caused by the COVID-19 pandemic, the number of foreign tourists variable is therefore directly added into the selected set of variables starting from 2019Q1 unless it is automatically selected by the LASSO.

Concerning for the candidate hyperparameters of the FA-LSTM model, we consider a following set of batch size = $\{\lceil{input/5}\rceil,\lceil{input/3}\rceil,\lceil{input/2}\rceil\}$; timesteps = $\{3,6,12,18\}$; and the number of LSTM memory cells = $\{(128),(256),(512),(128,128),(256,256)\}$, where the number of elements inside the brackets equals to the number of hidden layer(s). As for the lag specification, we consider a small set of monthly lag order = $\{3,6,9\}$ and quarterly lag order = $\{1,2,3\}$. The same candidate set of batch size, LSTM memory cells size, monthly lag order, and quarterly lag order are also considered for the FA-ANN model. As for the SA-LSTM model, the different set of candidate hyperparameters are considered except the batch size. We consider the larger set of the number of LSTM memory cells = $\{32,64,128,[32,32],[64,64],[128,128]\}$ and timesteps = $\{3,6,12,18,24\}$. Lastly, for the UNI-LSTM model, we consider a set of batch size = $\{\lceil{input/5}\rceil,\lceil{input/3}\rceil,\lceil{input/2}\rceil,\lceil{input/1}\rceil\}$; the number of LSTM memory cells = $\{(8),(16),(32),(64)\}$; and timesteps = $\{3,6,12,15,18,24\}$. To prevent overfitting, we employ the early stopping of five epochs and set the maximum epoch to 200. The optimal set of hyperparameters and lag specification is therefore selected from a total of 540, 90, 96, and 135 combinations when recursively performing the grid search for FA-LSTM, SA-LSTM, UNI-LSTM, and FA-ANN models, respectively.

\subsection{Empirical results}
Given the model configuration in each out-of-sample forecast for each forecasting horizon $h_m$, we conduct one-horizon-ahead forecast using the horizon-specific model. Except the simple AR(1) benchmark model, all competing models are estimated 100 times so as to account for the stochastic nature of the algorithm, and take the average value as the out-of-sample forecast of the respective model. The estimation sample depends on the model configuration and the information available at each period in time. 

Relative forecast accuracy of the competing models on the non-seasonally adjusted QGDP growth series can be found in Table \ref{table:empi}. We first focus only on the full out-of-sample 52 periods (2008Q1-2020Q4). The results show that the FA-LSTM model performs significantly better than the benchmark UNI-LSTM model only at the 6-month horizon, and is better, numerically but not statistically, at the 1-month horizon. The SA-LSTM model perform worse than those of the UNI-LSTM model at all horizons. The alternative FA-ANN model is better, numerically but not statistically, only at the 2-month horizon. Overall, the benchmark UNI-LSTM model seems to produce more accurate forecasts compared to these alternatives at all horizons. Compared with the simple benchmark AR(1) model, both FA-LSTM and the alternative FA-ANN models perform significantly better for all forecasting horizons. The SA-LSTM model also produce more accurate forecasts than the benchmark AR(1) for all horizon, the difference is only statistically significant at the 3-month horizon. Most of the relative MSFEs of all models are unsurprisingly less than one for all horizons as the variance of the non-seasonally adjusted QGDP growth series is considerably high. In terms of the relative MSFEs, the UNI-LSTM also provide the most information content at all horizons. Only the FA-LSTM model that is more informative at one and six months ahead. We also observe that, as new monthly information becomes available, the forecast accuracy of the FA-LSTM model at the nowcasting horizons ($h_m=1,2,3$) improve with the horizon. However, all competing models being considered here cannot always improve with this information. This could be a result of the non-optimality in our model configuration process that simultaneously select the optimal set of variables and hyperparameters.

\begin{table}[t!]
\centering
\begin{threeparttable}
\caption{Out-of-sample forecast comparisons for non-seasonally adjusted QGDP growth}
\label{table:empi}

\begin{tabular}{@{}rccccccccccc@{}}
\toprule
 & \multicolumn{3}{c}{2008Q1-2020Q4} &  & \multicolumn{3}{c}{Exclude large downturns} &  & \multicolumn{3}{c}{Large downturns} \\ \cmidrule(lr){2-4} \cmidrule(lr){6-8} \cmidrule(l){10-12} 
 & \multicolumn{2}{c}{Rel. RMSE} & \multirow{2}{*}{\begin{tabular}[c]{@{}c@{}}Rel.\\ MSE\end{tabular}} &  & \multicolumn{2}{c}{Rel. RMSE} & \multirow{2}{*}{\begin{tabular}[c]{@{}c@{}}Rel.\\ MSE\end{tabular}} &  & \multicolumn{2}{c}{Rel. RMSE} & \multirow{2}{*}{\begin{tabular}[c]{@{}c@{}}Rel.\\ MSE\end{tabular}} \\ \cmidrule(lr){2-3} \cmidrule(lr){6-7} \cmidrule(lr){10-11}
 & Uni & AR(1) &  &  & Uni & AR(1) &  &  & Uni & AR(1) &  \\ \cmidrule(lr){2-4} \cmidrule(lr){6-8} \cmidrule(l){10-12} 
\multicolumn{1}{l}{\textit{1-month horizon}} &  &  &  &  &  &  &  &  &  &  &  \\
FA-LSTM & 0.917 & \textbf{0.563} & 0.318 &  & \textbf{1.046} & \textbf{0.593} & 0.349 &  & \textbf{0.728} & \textbf{0.504} & 0.255 \\
SA-LSTM & 1.330 & 0.817 & 0.669 &  & 1.564 & \textbf{0.887} & 0.782 &  & 0.968 & \textbf{0.671} & 0.452 \\
UNI-LSTM & 3.593 & \textbf{0.615} & 0.378 &  & 3.206 & \textbf{0.567} & 0.320 & \textit{} & 4.340 & \textbf{0.693} & 0.482 \\
FA-ANN & \textbf{1.020} & \textbf{0.627} & 0.393 &  & 1.151 & \textbf{0.653} & 0.424 &  & 0.831 & \textbf{0.576} & 0.333 \\
AR(1) & \textbf{1.627} & 5.846 & 1.001 &  & \textbf{1.763} & 5.652 & 0.994 &  & \textbf{1.442} & 6.260 & 1.003 \\
\multicolumn{1}{l}{\textit{2-month horizon}} &  &  &  &  &  &  &  &  &  &  &  \\
FA-LSTM & \textbf{1.044} & \textbf{0.608} & 0.370 &  & \textbf{1.198} & \textbf{0.676} & 0.455 &  & \textbf{0.745} & \textbf{0.457} & 0.209 \\
SA-LSTM & \textbf{1.323} & 0.770 & 0.594 &  & \textbf{1.381} & \textbf{0.780} & 0.604 &  & \textbf{1.229} & \textbf{0.753} & 0.569 \\
UNI-LSTM & 3.403 & \textbf{0.582} & 0.339 &  & 3.191 & \textbf{0.565} & 0.317 &  & 3.837 & \textbf{0.613} & 0.377 \\
FA-ANN & 0.999 & \textbf{0.582} & 0.339 &  & \textbf{1.126} & \textbf{0.636} & 0.402 &  & \textbf{0.761} & \textbf{0.466} & 0.218 \\
AR(1) & \textbf{1.718} & 5.846 & 1.001 &  & \textbf{1.771} & 5.652 & 0.994 &  & \textbf{1.632} & 6.260 & 1.003 \\
\multicolumn{1}{l}{\textit{3-month horizon}} &  &  &  &  &  &  &  &  &  &  &  \\
FA-LSTM & \textbf{1.237} & \textbf{0.786} & 0.618 &  & \textbf{1.262} & \textbf{0.792} & 0.623 &  & \textbf{1.193} & \textbf{0.774} & 0.601 \\
SA-LSTM & \textbf{1.431} & \textbf{0.909} & 0.827 &  & \textbf{1.317} & \textbf{0.827} & 0.679 &  & \textbf{1.608} & \textbf{1.044} & 1.092 \\
UNI-LSTM & 3.713 & \textbf{0.635} & 0.404 &  & 3.547 & \textbf{0.628} & 0.391 &  & 4.062 & \textbf{0.649} & 0.422 \\
FA-ANN & 1.043 & \textbf{0.663} & 0.440 &  & \textbf{0.902} & \textbf{0.566} & 0.318 &  & \textbf{1.249} & \textbf{0.810} & 0.659 \\
AR(1) & \textbf{1.574} & 5.846 & 1.001 &  & \textbf{1.594} & 5.652 & 0.994 &  & \textbf{1.541} & 6.260 & 1.003 \\
\multicolumn{1}{l}{\textit{6-month horizon}} &  &  &  &  &  &  &  &  &  &  &  \\
FA-LSTM & \textbf{0.938} &  & 0.420 &  & \textbf{0.962} &  & 0.449 &  & \textbf{0.890} &  & 0.361 \\
SA-LSTM & \textbf{1.259} &  & 0.756 &  & \textbf{1.296} &  & 0.815 &  & \textbf{1.184} &  & 0.638 \\
UNI-LSTM & 4.034 &  & 0.477 &  & 3.950 &  & 0.485 &  & 4.217 &  & 0.455 \\
FA-ANN & \textbf{1.101} &  & 0.578 &  & \textbf{1.067} &  & 0.552 &  & \textbf{1.167} &  & 0.619 \\
\multicolumn{1}{l}{\textit{9-month horizon}} &  &  &  &  &  &  &  &  &  &  &  \\
FA-LSTM & \textbf{1.197} &  & 0.629 &  & 1.229 &  & 0.727 &  & 1.112 &  & 0.440 \\
SA-LSTM & \textbf{1.295} &  & 0.738 &  & \textbf{1.204} &  & 0.698 &  & \textbf{1.500} &  & 0.801 \\
UNI-LSTM & 3.874 &  & 0.440 &  & 3.936 &  & 0.482 &  & 3.731 &  & 0.356 \\
FA-ANN & 1.041 &  & 0.476 &  & \textbf{1.016} &  & 0.497 &  & \textbf{1.100} &  & 0.431 \\
\multicolumn{1}{l}{\textit{12-month horizon}} &  &  &  &  &  &  &  &  &  &  &  \\
FA-LSTM & 1.071 &  & 0.563 &  & \textbf{0.960} &  & 0.516 &  & \textbf{1.345} &  & 0.644 \\
SA-LSTM & \textbf{1.276} &  & 0.799 &  & \textbf{1.242} &  & 0.863 &  & \textbf{1.372} &  & 0.670 \\
UNI-LSTM & 4.092 &  & 0.490 &  & 4.243 &  & 0.560 &  & 3.728 &  & 0.356 \\
FA-ANN & 1.012 &  & 0.502 &  & \textbf{0.878} &  & 0.432 &  & \textbf{1.326} &  & 0.625 \\ \bottomrule
\end{tabular}

\begin{tablenotes}[flushleft]
      \small
      \item \textit{Notes}: This table reports out-of-sample forecast comparisons for six forecasting horizons ($h_m=1,2,3,6,9,12$). Columns 2-4 report comparison results for the full out-of-sample 52 periods (2008Q1-2020Q4). Columns 8-10 report comparison results for the large economic downturn periods (see main text for more discussion), while columns 5-7 report the results for the out-of-sample data which exclude the large economic downturn periods. In each comparison period, column \textit{Uni} and column \textit{AR(1)} report root mean squared forecasts error relative to the univariate LSTM and AR(1) models, respectively, while column \textit{Rel. MSE} reports mean squared forecasts error relative to the variance of QGDP growth. The variance of QGDP growth series in each evaluation period is 34.138, 32.151, and 39.066, respectively. As for the relative RMSFE measures, bold text indicates that the null hypothesis of the two methods having the same forecast accuracy is rejected at the 5\% significance level using \cite{diebold1995comparing} test statistics.
      
\end{tablenotes}
\end{threeparttable}
\end{table}

\begin{figure}[t!]
  \centering
  \includegraphics[scale=0.5]{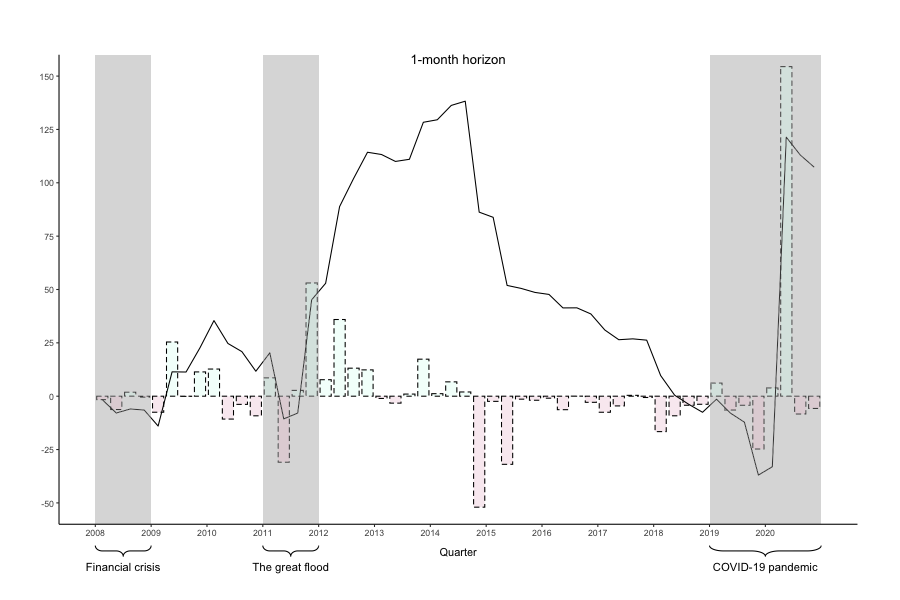}
  \caption{Cumulative sum of loss differentials (CUMSFE) of FA-LSTM model versus the benchmark univariate LSTM model at 1-month horizon.\\
  \textit{Notes}: Solid line corresponds to CUMSFE. Positive columns correspond to positive loss differential ($e_{q,uni}^2 - e_{q,fa}^2$) indicating that the FA-LSTM model has smaller squared forecast errors compared to the benchmark univariate LSTM model, and negative columns indicate the opposite.}
  \label{fig:fig3}
\end{figure}

Following \cite{babii2021machine}, we additionally examine whether the gains in predictive performance are persistent over recursions or simply dominated by differences in performance in only a few periods. We compute the cumulative sum of squared forecast error (CUMSFE) loss differential of the FA-LSTM and the benchmark UNI-LSTM models for the 1-month horizon. The CUMSFE for the benchmark UNI-LSTM model versus the FA-LSTM model for quarter $q = t,t+1,t+2,...,t+k$ is defined as:
\begin{equation}
CUMSFE_{t,t+k} = \sum_{q=t}^{t+k}(e_{q,uni}^2 - e_{q,fa}^2)
\end{equation}
where $t$ is the first prediction at the end of 2016Q1, $t+k$ is the last prediction at the end of 2020Q4, and $e_{q,uni}$ and $e_{q,fa}$ are the forecast errors for quarter $q$ from the benchmark UNI-LSTM and the FA-LSTM models, respectively. For an out-of-sample forecast of quarter $t$, a positive value of loss differential $e_{q,uni}^2 - e_{q,fa}^2$ indicates that the FA-LSTM model has smaller squared forecast errors compared to the benchmark UNI-LSTM model. Figure \ref{fig:fig3} shows the plot of CUMSFE and loss differential for the 1-month horizon. We observe that the gains in predictive performance are not persistent throughout the out-of-sample periods. Notably, the large gains (i.e. positive loss differential) of the FA-LSTM model can be observed during the unprecedented events like the 2011 great flood in Thailand and the COVID-19 pandemic. This result suggests that our predictive performance might be mostly driven by these unusual periods in our out-of-sample observations. To verify this, we thus turn our focus back to Table \ref{table:empi}. The last three columns (8-10) report the results for the out-of-sample forecasts in the financial crisis period, the great flood in Thailand period, and the COVID-19 pandemic period. For simplicity, we define the financial crisis period as 2008Q1-2008Q4; the great flood in Thailand period as 2011Q1-2011Q4, and the COVID-19 period as 2019Q1-2020Q4. Columns 5-7 report the results for the out-of-sample periods which exclude the aforementioned large economic downturns.

The results, in terms of ranking, for the out-of-sample forecasts in the periods of large economic downturn and the periods which exclude large economic downturn largely remain the same compared to the previously discussed ones. Here, we focus only on the nowcasting horizon $h_m=1,2,3$. The forecast accuracy of the FA-LSTM model improves in the large economic downturn periods; however, deteriorates when we exclude the large economic downturns. Conversely, the forecast accuracy of the benchmark UNI-LSTM model notably deteriorate in the large economic downturn, but improve when we remove the large economic downturn periods. Compared with the benchmark UNI-LSTM model, the results indicate that the FA-LSTM model perform significantly better during economic downturn periods at 1- and 2-month horizons; however, the benchmark UNI-LSTM model fits better when the economic downturn is absent. The results imply that our FA-LSTM model can help to nowcast QGDP during unprecedented events.

To be policy relevant, we additionally compare the annualized QGDP forecasts of the FA-LSTM model with the publicly available BOT model implied annual GDP forecasts at $h_m=1,3,6,9,12$. To this end, we exploit both the forecasts produced by FA-LSTM model and the actual QGDP of the prior quarter(s) in the same year. For the 12-month ahead ($h_m=12$) annual forecast at the beginning of 2008, for example, we use the forecasts of 2008Q1 at 3-month horizon, 2008Q2 at 6-month horizon, 2008Q3 at 9-month horizon, and 2008Q4 at 12-month horizon. Then, for the 9-month ahead ($h_m=9$) annual forecast at the beginning of April 2008, we use the actual QGDP growth of 2008Q1, the forecasts of 2008Q2 at 3-month horizon, 2008Q3 at 6-month horizon, and 2008Q4 at 9-month horizon. Note that we abstract from additional complications such as those resulting from publication delay of the QGDP series.

\begin{figure}[tbh!]
  \centering
  \includegraphics[width=1\textwidth]{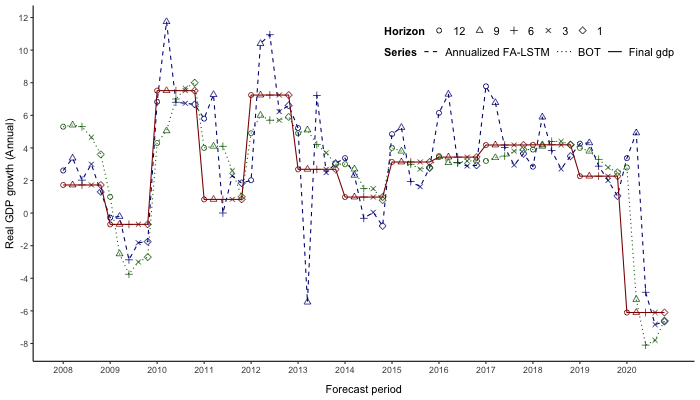}
  \caption{Annualized QGDP growth of the FA-LSTM model versus the BOT model implied annual GDP growth. \\
  \textit{Notes}: For each year, the projection of annual GDP growth are made at the beginning of January ($h_m=12$), April ($h_m=9$), July ($h_m=6$), October ($h_m=3$), and December ($h_m=1$). Dashed line corresponds to annualized QGDP growth produced by the FA-LSTM model, dotted line corresponds to the BOT model implied annual GDP growth, and solid line corresponds to the final annual GDP growth.}
  \label{fig:fig4}
\end{figure}

Concerning for the BOT forecast, the figures are compiled from various documents officially published by the BOT. Specifically, we take the annual GDP growth forecasts that appear in the minutes of the monetary policy committee meetings, the monetary policy report, inflation report, and the press release, which are all publicly available on the BOT websites. We use the first available forecast that published later than the beginning of the annualized QGDP forecast period for each forecasting horizon. Again, for example, for the 12-month ahead ($h_m=12$) annual forecast at the beginning of 2008, the 2008 annual GDP growth forecast published in January 2008 inflation report is used.

In Figure \ref{fig:fig4}, we plot the annualized QGDP growth forecast of the FA-LSTM model versus BOT model implied annual forecast for $h_m=1,3,6,9,12$. In each year, we intentionally plot all five forecasts to present the development of the forecasts in each forecasting horizon $h_m$. The BOT model unsurprisingly produces smoother forecasts as the FA-LSTM model performs poorly at longer horizons, i.e. $h_m=9,12$. However, we can observe substantially improvement at the shorter horizon $h_m=1$ and $3$. In Table \ref{table:annualbot}, we reports the predictive performance of both models at each forecasting horizon. The results confirm that the forecast accuracy of the annualized QGDP growth forecast of the FA-LSTM model is better than the BOT model implied forecast at shorter horizons $h_m=1$ and $3$. While the BOT model implied forecasts have better forecast accuracy than the annualized QGDP growth forecasts at longer horizons $h_m=9$ and $12$; the forecast accuracy of both models at 6-month horizon are not substantially different.

\begin{table}[t!]
\centering
\begin{threeparttable}
\caption{Forecast comparisons for annual GDP growth}
\label{table:annualbot}

\begin{tabular}{@{}rccc@{}}
\toprule
                 & \multicolumn{3}{c}{RMSE}      \\ \cmidrule(l){2-4} 
                 & Annualized FA-LSTM &  & BOT   \\ \cmidrule(lr){2-2} \cmidrule(l){4-4} 
1-month horizon  & 0.861              &  & 0.890 \\
3-month horizon  & 1.054              &  & 1.367 \\
6-month horizon  & 1.873              &  & 1.833 \\
9-month horizon  & 4.764              &  & 1.920 \\
12-month horizon & 3.764              &  & 3.215 \\ \bottomrule
\end{tabular}

\begin{tablenotes}[flushleft]
      \small
      \item \textit{Notes}: This table reports out-of-sample forecast comparisons for five forecasting horizons ($h_m=1,3,6,9,12$). The last two columns report the RMSFE of the forecasts between 2008-2020 based on the annualized QGDP growth of the FA-LSTM model and BOT model implied annual GDP growth forecasts, respectively.
      
\end{tablenotes}
\end{threeparttable}
\end{table}

\section{Conclusions}
\label{sec:others}
This paper demonstrates the potentials of applying the LSTM with time series data sampled at different frequencies. Our contribution is to combine methods from the literature on machine learning and on mixed frequency time series data. We adopt the U-MIDAS scheme into the LSTM architecture as a forecasting tool. From machine learning perspective, the U-MIDAS scheme serves as an alternative feature engineering process that transforms the higher-frequency variables into a low-frequency vector, denoted frequency alignment, and allows us to exploit potentially useful information contained in the high-frequency time series data.

We have shown via Monte Carlo simulation the predictive performance of the LSTM with sampling alignment strategy (i.e. the SA-LSTM), and the LSTM using U-MIDAS scheme (i.e. the FA-LSTM). Both models largely outperform the MIDAS regression even in a set up favorable to the restricted MIDAS. The simulation results suggest that the FA-LSTM is particularly suited to provide nowcasts of output variables in the quarterly/monthly data with small sample size. We then continue our assessment in the empirical application for quarterly growth rate of Thai real GDP. The results show that the univariate LSTM is generally a strong benchmark compared to the commonly used AR(1) model. The SA-LSTM performs poorly compared to the benchmark univariate LSTM model at all horizons. We find that the FA-LSTM performs significantly better than the benchmark univariate LSTM model only at the 6-month horizon, and is better, numerically but not statistically, at the 1-month horizon. To be policy relevant, we additionally compare the annualized QGDP forecasts of the FA-LSTM with the publicly available BOT model implied annual GDP forecasts for 1-, 3-, 6-, 9-, and 12-month horizons. The results show that the forecast accuracy of the annualized QGDP growth forecast of the FA-LSTM model is better than the BOT model implied forecast at shorter horizons ($h_m=1,3$).

To sum up, both the Monte Carlo simulation and the empirical application show that the LSTM using U-MIDAS scheme provides more flexibility when ones need to nowcast the macroeconomic low-frequency data series, such as GDP growth, in the quarterly/monthly data. This does not mean that our proposed LSTM using U-MIDAS scheme can not be applied to other frequency mismatch ratios. The frequency alignment procedure can be generalized to other data sampled at different frequencies, such as daily or unconventional search data. Nonetheless, there are some limitations that should be taken into account for future research. Similar to conventional LSTM model, the proposed LSTM using U-MIDAS scheme requires an optimal set of hyperparameters. And even worse, the proposed model additionally requires an optimal set of lag specification; consequently, the computational burden is even larger as a large number of hyperparameters and lag specification combinations are required to be considered. The model configuration strategy used in this paper is very time-consuming and obviously not optimal. This is an interesting area for future research.

\bibliography{references.bib}

\begin{thebibliography}{}

\bibitem[\protect\citeauthoryear{Andreou, Ghysels \& Kourtellos}{Andreou
  et~al.}{2013}]{andreou2013should}
Andreou, E., Ghysels, E., \& Kourtellos, A. (2013).
\newblock Should macroeconomic forecasters use daily financial data and how?
\newblock {\em Journal of Business \& Economic Statistics}, {\em 31\/}(2),
  240--251.

\bibitem[\protect\citeauthoryear{Armesto, Hern{\'a}ndez-Murillo, Owyang \&
  Piger}{Armesto et~al.}{2009}]{armesto2009measuring}
Armesto, M.~T., Hern{\'a}ndez-Murillo, R., Owyang, M.~T., \& Piger, J. (2009).
\newblock Measuring the information content of the beige book: A mixed data
  sampling approach.
\newblock {\em Journal of Money, Credit and Banking}, {\em 41\/}(1), 35--55.

\bibitem[\protect\citeauthoryear{Babii, Ghysels \& Striaukas}{Babii
  et~al.}{2021}]{babii2021machine}
Babii, A., Ghysels, E., \& Striaukas, J. (2021).
\newblock Machine learning time series regressions with an application to
  nowcasting.

\bibitem[\protect\citeauthoryear{Breitung \& Roling}{Breitung \&
  Roling}{2015}]{breitung2015forecasting}
Breitung, J. \& Roling, C. (2015).
\newblock Forecasting inflation rates using daily data: A nonparametric midas
  approach.
\newblock {\em Journal of Forecasting}, {\em 34\/}(7), 588--603.

\bibitem[\protect\citeauthoryear{Choi \& Varian}{Choi \&
  Varian}{2012}]{choi2012predicting}
Choi, H. \& Varian, H. (2012).
\newblock Predicting the present with google trends.
\newblock {\em Economic record}, {\em 88}, 2--9.

\bibitem[\protect\citeauthoryear{Clements \& Galv{\~a}o}{Clements \&
  Galv{\~a}o}{2008}]{clements2008macroeconomic}
Clements, M.~P. \& Galv{\~a}o, A.~B. (2008).
\newblock Macroeconomic forecasting with mixed-frequency data: Forecasting
  output growth in the united states.
\newblock {\em Journal of Business \& Economic Statistics}, {\em 26\/}(4),
  546--554.

\bibitem[\protect\citeauthoryear{Diebold \& Mariano}{Diebold \&
  Mariano}{1995}]{diebold1995comparing}
Diebold, F.~X. \& Mariano, R.~S. (1995).
\newblock Comparing predictive accuracy.
\newblock {\em Journal of Business \& Economic Statistics}, {\em 13\/}(3),
  253--263.

\bibitem[\protect\citeauthoryear{D’Amuri \& Marcucci}{D’Amuri \&
  Marcucci}{2017}]{d2017predictive}
D’Amuri, F. \& Marcucci, J. (2017).
\newblock The predictive power of google searches in forecasting us
  unemployment.
\newblock {\em International Journal of Forecasting}, {\em 33\/}(4), 801--816.

\bibitem[\protect\citeauthoryear{Ettredge, Gerdes \& Karuga}{Ettredge
  et~al.}{2005}]{ettredge2005using}
Ettredge, M., Gerdes, J., \& Karuga, G. (2005).
\newblock Using web-based search data to predict macroeconomic statistics.
\newblock {\em Communications of the ACM}, {\em 48\/}(11), 87--92.

\bibitem[\protect\citeauthoryear{Fischer \& Krauss}{Fischer \&
  Krauss}{2018}]{fischer2018deep}
Fischer, T. \& Krauss, C. (2018).
\newblock Deep learning with long short-term memory networks for financial
  market predictions.
\newblock {\em European Journal of Operational Research}, {\em 270\/}(2),
  654--669.

\bibitem[\protect\citeauthoryear{Foroni \& Marcellino}{Foroni \&
  Marcellino}{2013}]{foroni2013survey}
Foroni, C. \& Marcellino, M. (2013).
\newblock A survey of econometric methods for mixed-frequency data.
\newblock Technical report, European University Institute.

\bibitem[\protect\citeauthoryear{Foroni, Marcellino \& Schumacher}{Foroni
  et~al.}{2015}]{foroni2015unrestricted}
Foroni, C., Marcellino, M., \& Schumacher, C. (2015).
\newblock Unrestricted mixed data sampling (midas): Midas regressions with
  unrestricted lag polynomials.
\newblock {\em Journal of the Royal Statistical Society: Series A (Statistics
  in Society)}, {\em 178\/}(1), 57--82.

\bibitem[\protect\citeauthoryear{Gers, Schmidhuber \& Cummins}{Gers
  et~al.}{2000}]{gers2000learning}
Gers, F.~A., Schmidhuber, J., \& Cummins, F. (2000).
\newblock Learning to forget: Continual prediction with lstm.
\newblock {\em Neural computation}, {\em 12\/}(10), 2451--2471.

\bibitem[\protect\citeauthoryear{Ghysels, Horan \& Moench}{Ghysels
  et~al.}{2018}]{ghysels2018forecasting}
Ghysels, E., Horan, C., \& Moench, E. (2018).
\newblock Forecasting through the rearview mirror: Data revisions and bond
  return predictability.
\newblock {\em The Review of Financial Studies}, {\em 31\/}(2), 678--714.

\bibitem[\protect\citeauthoryear{Ghysels, Kvedaras \& Zemlys}{Ghysels
  et~al.}{2016}]{ghysels2016mixed}
Ghysels, E., Kvedaras, V., \& Zemlys, V. (2016).
\newblock Mixed frequency data sampling regression models: the r package
  midasr.
\newblock {\em Journal of statistical software}, {\em 72\/}(1), 1--35.

\bibitem[\protect\citeauthoryear{Ghysels, Santa-Clara \& Valkanov}{Ghysels
  et~al.}{2004}]{ghysels2004midas}
Ghysels, E., Santa-Clara, P., \& Valkanov, R. (2004).
\newblock The midas touch: Mixed data sampling regression models.
\newblock Technical report, Anderson Graduate School of Management, UCLA.

\bibitem[\protect\citeauthoryear{Ghysels, Santa-Clara \& Valkanov}{Ghysels
  et~al.}{2005}]{ghysels2005there}
Ghysels, E., Santa-Clara, P., \& Valkanov, R. (2005).
\newblock There is a risk-return trade-off after all.
\newblock {\em Journal of Financial Economics}, {\em 76\/}(3), 509--548.

\bibitem[\protect\citeauthoryear{Ghysels, Santa-Clara \& Valkanov}{Ghysels
  et~al.}{2006}]{ghysels2006predicting}
Ghysels, E., Santa-Clara, P., \& Valkanov, R. (2006).
\newblock Predicting volatility: getting the most out of return data sampled at
  different frequencies.
\newblock {\em Journal of Econometrics}, {\em 131\/}(1-2), 59--95.

\bibitem[\protect\citeauthoryear{Ghysels, Sinko \& Valkanov}{Ghysels
  et~al.}{2007}]{ghysels2007midas}
Ghysels, E., Sinko, A., \& Valkanov, R. (2007).
\newblock Midas regressions: Further results and new directions.
\newblock {\em Econometric Reviews}, {\em 26\/}(1), 53--90.

\bibitem[\protect\citeauthoryear{Giannone, Reichlin \& Small}{Giannone
  et~al.}{2008}]{giannone2008nowcasting}
Giannone, D., Reichlin, L., \& Small, D. (2008).
\newblock Nowcasting: The real-time informational content of macroeconomic
  data.
\newblock {\em Journal of Monetary Economics}, {\em 55\/}(4), 665--676.

\bibitem[\protect\citeauthoryear{Graves}{Graves}{2013}]{graves2013generating}
Graves, A. (2013).
\newblock Generating sequences with recurrent neural networks.

\bibitem[\protect\citeauthoryear{Hochreiter, Bengio, Frasconi, Schmidhuber \&
  others}{Hochreiter et~al.}{2001}]{hochreiter2001gradient}
Hochreiter, S., Bengio, Y., Frasconi, P., Schmidhuber, J., et~al. (2001).
\newblock Gradient flow in recurrent nets: the difficulty of learning long-term
  dependencies.

\bibitem[\protect\citeauthoryear{Hochreiter \& Schmidhuber}{Hochreiter \&
  Schmidhuber}{1997}]{hochreiter1997long}
Hochreiter, S. \& Schmidhuber, J. (1997).
\newblock Long short-term memory.
\newblock {\em Neural computation}, {\em 9\/}(8), 1735--1780.

\bibitem[\protect\citeauthoryear{Koenig, Dolmas \& Piger}{Koenig
  et~al.}{2003}]{koenig2003use}
Koenig, E.~F., Dolmas, S., \& Piger, J. (2003).
\newblock The use and abuse of real-time data in economic forecasting.
\newblock {\em Review of Economics and Statistics}, {\em 85\/}(3), 618--628.

\bibitem[\protect\citeauthoryear{Kuzin, Marcellino \& Schumacher}{Kuzin
  et~al.}{2011}]{kuzin2011midas}
Kuzin, V., Marcellino, M., \& Schumacher, C. (2011).
\newblock Midas vs. mixed-frequency var: Nowcasting gdp in the euro area.
\newblock {\em International Journal of Forecasting}, {\em 27\/}(2), 529--542.

\bibitem[\protect\citeauthoryear{Marcellino \& Schumacher}{Marcellino \&
  Schumacher}{2010}]{marcellino2010factor}
Marcellino, M. \& Schumacher, C. (2010).
\newblock Factor midas for nowcasting and forecasting with ragged-edge data: A
  model comparison for german gdp.
\newblock {\em Oxford Bulletin of Economics and Statistics}, {\em 72\/}(4),
  518--550.

\bibitem[\protect\citeauthoryear{Qiu}{Qiu}{2020}]{qiu2020forecasting}
Qiu, Y. (2020).
\newblock Forecasting the consumer confidence index with tree-based midas
  regressions.
\newblock {\em Economic Modelling}, {\em 91}, 247--256.

\bibitem[\protect\citeauthoryear{Sax \& Eddelbuettel}{Sax \&
  Eddelbuettel}{2018}]{sax2018seasonal}
Sax, C. \& Eddelbuettel, D. (2018).
\newblock Seasonal adjustment by x-13arima-seats in r.
\newblock {\em Journal of Statistical Software}, {\em 87\/}(1), 1--17.

\bibitem[\protect\citeauthoryear{Sezer, Gudelek \& Ozbayoglu}{Sezer
  et~al.}{2020}]{sezer2020financial}
Sezer, O.~B., Gudelek, M.~U., \& Ozbayoglu, A.~M. (2020).
\newblock Financial time series forecasting with deep learning: A systematic
  literature review: 2005--2019.
\newblock {\em Applied Soft Computing}, {\em 90}, 106181.

\bibitem[\protect\citeauthoryear{Tibshirani}{Tibshirani}{1996}]{tibshirani1996regression}
Tibshirani, R. (1996).
\newblock Regression shrinkage and selection via the lasso.
\newblock {\em Journal of the Royal Statistical Society: Series B
  (Methodological)}, {\em 58\/}(1), 267--288.

\bibitem[\protect\citeauthoryear{Vosen \& Schmidt}{Vosen \&
  Schmidt}{2011}]{vosen2011forecasting}
Vosen, S. \& Schmidt, T. (2011).
\newblock Forecasting private consumption: survey-based indicators vs. google
  trends.
\newblock {\em Journal of forecasting}, {\em 30\/}(6), 565--578.

\bibitem[\protect\citeauthoryear{Xu, Zhuo, Jiang \& Liu}{Xu
  et~al.}{2019}]{xu2019artificial}
Xu, Q., Zhuo, X., Jiang, C., \& Liu, Y. (2019).
\newblock An artificial neural network for mixed frequency data.
\newblock {\em Expert Systems with Applications}, {\em 118}, 127--139.

\end{thebibliography}

\newpage

\appendix
\setcounter{table}{0}
\renewcommand{\thetable}{\Alph{section}.\arabic{table}}
\renewcommand{\appendixpagename}{\centering APPENDIX}
\appendixpage
\renewcommand{\thesection}{Appendix \Alph{section}}

\section{Optimal Hyperparameters selected for Monte Carlo Simulation}

\begin{table}[h!]
\caption{Optimal hyperparameter combinations for Monte Carlo simulation}
\label{table:mc_para}
\centering
\begin{threeparttable}
\tiny
\begin{tabular}{@{}ccccccccccccccccccccccc@{}}
\toprule
\multirow{3}{*}{$h_m$} & \multirow{3}{*}{\begin{tabular}[c]{@{}c@{}}Raw\\ Obs\end{tabular}} & \multicolumn{8}{c}{SA-LSTM} &  & \multicolumn{12}{c}{FA-LSTM} \\ 
\cmidrule(lr){3-10} \cmidrule(l){12-23} &  & \multicolumn{4}{c}{{[}6, 0:0{]}} & \multicolumn{4}{c}{{[}12, 0:0{]}} &  & \multicolumn{4}{c}{{[}4, 0:2{]}} & \multicolumn{4}{c}{{[}2, 0:5{]}} & \multicolumn{4}{c}{{[}1, 0:11{]}} \\ \cmidrule(lr){3-10} \cmidrule(l){12-23} 
&  & p1 & p2 & p3 & p4 & p1 & p2 & p3 & p4 &  & p1 & p2 & p3 & p4 & p1 & p2 & p3 & p4 & p1 & p2 & p3 & p4 \\ \midrule \\
   &    &    &    &    &    & \multicolumn{13}{c}{Panel A. $x_{k,\tau} \sim N(0,1)$}  &    &.   &    &    \\ \\
1  & 50 & 25 & 0.4 & 15 & 32 & 25 & 0.4 & 15 & 32 &  & 25 & 0.4 & 14 & 16 & 25 & 0.4 & 15 & 32 & 25 & 0.0 & 15 & 16 \\
   & 80 & 50 & 0.4 & 3  & 8  & 25 & 0.4 & 5  & 16 &  & 25 & 0.4 & 23 & 32 & 25 & 0.0 & 24 & 32 & 25 & 0.4 & 24 & 16 \\
2  & 50 & 25 & 0.4 & 15 & 16 & 50 & 0.0 & 15 & 8  &  & 25 & 0.0 & 14 & 16 & 25 & 0.0 & 15 & 32 & 25 & 0.0 & 15 & 32 \\
   & 80 & 50 & 0.4 & 24 & 16 & 25 & 0.4 & 24 & 32 &  & 25 & 0.4 & 23 & 32 & 25 & 0.0 & 24 & 16 & 25 & 0.0 & 24 & 16 \\
3  & 50 & 50 & 0.4 & 15 & 8  & 25 & 0.4 & 3  & 8  &  & 25 & 0.4 & 14 & 32 & 25 & 0.4 & 15 & 32 & 25 & 0.4 & 15 & 64 \\
   & 80 & 25 & 0.4 & 24 & 32 & 25 & 0.0 & 24 & 32 &  & 25 & 0.0 & 23 & 32 & 25 & 0.4 & 24 & 16 & 25 & 0.4 & 24 & 32 \\
6  & 50 & 50 & 0.0 & 15 & 8  & 25 & 0.4 & 3  & 8  &  & 25 & 0.0 & 14 & 16 & 25 & 0.4 & 15 & 32 & 25 & 0.0 & 14 & 32 \\
   & 80 & 25 & 0.4 & 24 & 32 & 25 & 0.4 & 5  & 16 &  & 25 & 0.4 & 23 & 32 & 25 & 0.0 & 24 & 32 & 25 & 0.4 & 23 & 32 \\
9  & 50 & 25 & 0.0 & 14 & 8  & 50 & 0.4 & 14 & 8  &  & 25 & 0.0 & 14 & 16 & 25 & 0.0 & 14 & 16 & 25 & 0.4 & 14 & 16 \\
   & 80 & 25 & 0.4 & 23 & 32 & 25 & 0.4 & 23 & 32 &  & 25 & 0.0 & 23 & 16 & 25 & 0.4 & 23 & 16 & 25 & 0.0 & 23 & 16 \\
12 & 50 & 25 & 0.4 & 13 & 16 & 50 & 0.4 & 13 & 16 &  & 25 & 0.4 & 13 & 16 & 25 & 0.0 & 14 & 16 & 25 & 0.0 & 13 & 16 \\
   & 80 & 25 & 0.0 & 22 & 16 & 25 & 0.4 & 22 & 32 &  & 25 & 0.0 & 22 & 32 & 25 & 0.0 & 23 & 16 & 25 & 0.0 & 22 & 16 \\ \midrule \\
   &    &    &     &    &    & \multicolumn{13}{c}{Panel B. $x_{k,\tau} = 0.9 x_{k,\tau-1}+\epsilon_{\tau}$} 
   &    &    &     &    \\ \\
1  & 50 & 50 & 0.4 & 3  & 16 & 50 & 0.4 & 15 & 32 &  & 25 & 0.4 & 1  & 128 & 25 & 0.4 & 15 & 128 & 25 & 0.0 & 1  & 128 \\
   & 80 & 50 & 0.0 & 3  & 16 & 25 & 0.4 & 1  & 32 &  & 25 & 0.0 & 23 & 64  & 50 & 0.4 & 24 & 64  & 50 & 0.4 & 24 & 32  \\
2  & 50 & 50 & 0.4 & 1  & 8  & 50 & 0.4 & 3  & 8  &  & 50 & 0.4 & 1  & 128 & 25 & 0.4 & 15 & 16  & 50 & 0.0 & 1  & 64  \\
   & 80 & 50 & 0.0 & 5  & 16 & 50 & 0.4 & 5  & 16 &  & 50 & 0.4 & 23 & 16  & 25 & 0.4 & 24 & 128 & 25 & 0.4 & 24 & 128 \\
3  & 50 & 50 & 0.4 & 15 & 16 & 25 & 0.4 & 1  & 8  &  & 25 & 0.0 & 14 & 16  & 25 & 0.0 & 15 & 32  & 25 & 0.4 & 15 & 32  \\
   & 80 & 25 & 0.4 & 5  & 32 & 50 & 0.4 & 24 & 32 &  & 25 & 0.0 & 23 & 64  & 25 & 0.4 & 24 & 64  & 25 & 0.0 & 24 & 128 \\
6  & 50 & 50 & 0.4 & 15 & 32 & 25 & 0.4 & 14 & 32 &  & 25 & 0.4 & 14 & 16  & 25 & 0.0 & 15 & 32  & 25 & 0.4 & 14 & 32  \\
   & 80 & 25 & 0.0 & 24 & 32 & 50 & 0.4 & 23 & 16 &  & 25 & 0.4 & 5  & 16  & 25 & 0.4 & 24 & 32  & 25 & 0.0 & 23 & 16  \\
9  & 50 & 50 & 0.4 & 14 & 8  & 25 & 0.4 & 3  & 32 &  & 25 & 0.4 & 14 & 16  & 25 & 0.0 & 14 & 32  & 25 & 0.4 & 14 & 32  \\
   & 80 & 25 & 0.4 & 5  & 16 & 25 & 0.4 & 1  & 32 &  & 25 & 0.4 & 23 & 16  & 25 & 0.4 & 23 & 32  & 25 & 0.4 & 23 & 16  \\
12 & 50 & 50 & 0.4 & 13 & 8  & 25 & 0.0 & 3  & 8  &  & 25 & 0.0 & 13 & 16  & 25 & 0.4 & 14 & 16  & 25 & 0.0 & 13 & 32  \\
   & 80 & 25 & 0.0 & 22 & 32 & 50 & 0.4 & 22 & 8  &  & 25 & 0.4 & 22 & 16  & 25 & 0.4 & 23 & 16  & 25 & 0.0 & 22 & 16  \\ \bottomrule
\end{tabular}
\end{threeparttable}
\end{table}

This table reports the optimal set of hyperparameters selected by the grid search method discussed in the main text.  Column 1 and 2 indicate the higher-frequency forecasting horizons and the sample size, respectively. The remaining columns respectively report the optimal values of epoch ($p_1$), dropout ($p_2$), batch size ($p_3$), and the number of LSTM memory cells ($p_4$) for the LSTM models in each forecasting horizon. Panel A and B report the optimal hyperparameters for the experiments in which the higher-frequency variables are generated as $x_{k,\tau} \sim N(0,1)$ and AR(1) process with persistence equal to 0.9, respectively. The timesteps and lag specification of the input data, which are fixed across all forecasting horizons, are shown for each LSTM model. Since the SA-LSTM models take no lagged variables, the lag specification of [0:0] is simply shown. Given that three higher-frequency variables ($K = 3$) are used, the total number of variables used for each LSTM model are thus $K \times (J+1) = 3,3,9,18,36$.

\newpage
\setcounter{table}{0}
\section{Data description}
All quarterly and monthly macroeconomic series are from the \href{https://www.bot.or.th/English/Statistics/Pages/default.aspx}{Bank of Thailand (BOT)} database. All series are final revised data and supposedly published with a delay of up to one month according to the observed typical pattern between the end of the reference period and the date of the respective release.

\begin{table}[h!]
\caption{Data description table}
\label{table:data}
\centering
\begin{threeparttable}
\begin{tabular}{@{}lllll@{}}
\toprule
Description & Identifier & Since & SA & Transformation \\ \midrule
1 year treasury Bill & FM\_RT\_001 & 2001M1 & No & MoM \% change \\
10 years treasury bill & FM\_RT\_001 & 2001M1 & No & MoM \% change \\
USD Mid FX RATE & FM\_FX\_001 & 1981M1 & No & MoM \% change \\
GBP Mid FX RATE & FM\_FX\_001 & 1981M1 & No & MoM \% change \\
JPY Mid FX RATE & FM\_FX\_001 & 1981M1 & No & MoM \% change \\
Manufacturing production index & EC\_EI\_001 & 1993M1 & Yes & MoM \% change \\
Gross value added tax (at 2000 prices) & EC\_EI\_001 & 1993M1 & Yes & MoM \% change \\
Domestic automobiles sales & EC\_EI\_001 & 1993M1 & Yes & MoM \% change \\
Authorized capital of newly registered companies & EC\_EI\_002 & 1993M1 & Yes & MoM \% change \\
Construction areas permitted & EC\_EI\_002 & 1993M1 & Yes & MoM \% change \\
Number of foreign tourists & EC\_EI\_028 & 1993M1 & Yes & Level change \\
Stock exchange of Thailand (SET) index & EC\_EI\_002 & 1993M1 & Yes & MoM \% change \\
Broad money (at 2000 prices) & EC\_EI\_002 & 1993M1 & Yes & MoM \% change \\
Oil price inverse index & EC\_EI\_002 & 1993M1 & Yes & MoM \% change \\
Business sentiment index - performance & EC\_EI\_005 & 1999M1 & No & Index \\
Business sentiment index - total order book & EC\_EI\_005 & 1999M1 & No & Index \\
Business sentiment index - investment & EC\_EI\_005 & 1999M1 & No & Index \\
Business sentiment index - employment & EC\_EI\_005 & 1999M1 & No & Index \\
Land and building transactions nationwide & EC\_EI\_009 & 1995M1 & No & MoM \% change \\
Registered applicants & EC\_EI\_018 & 1995M1 & No & MoM \% change \\
Vacancies & EC\_EI\_018 & 1995M1 & No & MoM \% change \\
Placements & EC\_EI\_018 & 1995M1 & No & MoM \% change \\
Retail sales index - non-durable goods & EC\_EI\_032 & 2000M1 & Yes & MoM \% change \\
Retail sales index - durable goods & EC\_EI\_032 & 2000M1 & Yes & MoM \% change \\
Retail sales index - department stores, etc. & EC\_EI\_032 & 2000M1 & Yes & MoM \% change \\
Retail sales index - motor vehicles and fuel & EC\_EI\_032 & 2000M1 & Yes & MoM \% change \\
Wholesales index - non-durable goods & EC\_EI\_033 & 2000M1 & Yes & MoM \% change \\
Wholesales index - durable goods & EC\_EI\_033 & 2000M1 & Yes & MoM \% change \\
Wholesales index - intermediate goods & EC\_EI\_033 & 2000M1 & Yes & MoM \% change \\
Agricultural products export value index & EC\_EI\_026 & 2000M1 & Yes & MoM \% change \\
Fishery products export value index & EC\_EI\_026 & 2000M1 & Yes & MoM \% change \\
Manufactured products export value index & EC\_EI\_026 & 2000M1 & Yes & MoM \% change \\
Consumer goods import value index & EC\_EI\_025 & 2000M1 & Yes & MoM \% change \\
Raw materials import value index & EC\_EI\_025 & 2000M1 & Yes & MoM \% change \\
Capital goods import value index & EC\_EI\_025 & 2000M1 & Yes & MoM \% change \\
Land index & EC\_EI\_008 & 1991Q1 & No & MoM \% change \\
Single-detached house (including land) index & EC\_EI\_008 & 1991Q1 & No & MoM \% change \\
Town house (including land) index & EC\_EI\_008 & 1991Q1 & No & MoM \% change \\
Total gross NPLs outstanding & FI\_NP\_003 & 1991Q1 & No & MoM \% change \\
Total credit outstanding of credit card & FI\_CB\_080 & 1991Q1 & Yes & MoM \% change \\ \bottomrule
\end{tabular}
\begin{tablenotes}[flushleft]
      \item \textit{Notes}: The \textit{Description} column presents a macroeconomic time series name. The second column \textit{Identifier} denotes an assigned Report ID of a time series in the BOT database. The third column gives the first available period of information for a time series. The column \textit{SA} reports whether a seasonal patterns of a time series is removed. The last column \textit{Transformation} denotes the data transformation applied to a time series.
\end{tablenotes}
\end{threeparttable}
\end{table}

Table \ref{table:data} reports the transformation applied to each macroeconomic time series. Most of the macroeconomic series are transformed using their original information. The following macroeconomic series are needed to be processed prior to subsequent transformation: the monthly number of foreign tourists, the monthly business sentiment indexes, the monthly retail sales indexes, and the monthly wholesales indexes.

To this end, we re-reference all four series of the monthly business sentiment index from 50 to zero so as to clearly represent the direction of business sentiment. The seasonal patterns of all series of the monthly retail sales indexes and the monthly wholesales indexes are removed using the X-13ARIMA-SEATS algorithm available in the seasonal package \citep{sax2018seasonal} in R prior to calculating the percentage change. The monthly number of foreign tourists data used in this paper are from two series. The old series of the seasonally adjusted number of foreign tourists had been published since 1993M1 and was obsolete after 1996M12. The BOT has since published the new series of non-seasonally adjusted number of foreign tourists until present. To this end, we first deseasonalized the number of foreign tourists from 1997M1 to 20203 using the X-13ARIMA-SEATS algorithm. The observations from 2020M4 to 2020M12 are intentionally excluded to avoid inappropriate deseasonalization because the number of foreign tourists are either zero or drastically low since 2020M4 due to the COVID-19 pandemic. The original data are simply used if the observations from 2020M4 to 2020M12 are required when estimating the model.

\end{document}